\documentclass[useAMS,usenatbib]{mn2e}
\usepackage{graphicx}
\usepackage{textcomp}
\usepackage{textgreek}
\usepackage[T1]{fontenc}
\usepackage{amssymb,amsmath,enumitem}
\usepackage{hyperref}
\usepackage{breakurl}
\usepackage{booktabs}
\newcommand{\Angst}{\textup{\AA}}

\newcommand{\lnh}{\hbox{log~$N_{\rm H}$}}

\newcommand{\lLa}{\hbox{log~$L_{2-10}$}}

\newcommand{\kms}{\hbox{km~s$^{-1}$}}
\newcommand{\cmsq}{\hbox{cm$^{-2}$}}
\newcommand{\cc}{\hbox{cm$^{-3}$}}
\newcommand{\flux}{\hbox{erg~cm$^{-2}$~s$^{-1}$}}
\newcommand{\lumin}{\hbox{erg~s$^{-1}$}}

\newcommand{\aox}{\hbox{$\alpha_{\rm ox}$}}
\newcommand{\aeuv}{\hbox{$\alpha_{EUV}$}}
\newcommand{\afuv}{\hbox{$\alpha_{FUV}$}}
\newcommand{\nh}{\hbox{${N}_{\rm H}$}}

\newcommand{\den}{\hbox{$n_{\rm H}$}}

\newcommand{\pg}{\hbox{PG~2112+059}}
\newcommand{\RF}{\hbox{rest-frame}}

\newcommand{\FeXXV}{\hbox{Fe {\sc xxv}}}
\newcommand{\FeXXVI}{\hbox{Fe {\sc xxvi}}}
\newcommand{\AlIII}{\hbox{Al {\sc iii}}}

\newcommand{\CIV}{\hbox{C {\sc iv}}}
\newcommand{\SiIV}{\hbox{Si {\sc iv}}}
\newcommand{\SiIII}{\hbox{Si {\sc iii}}}
\newcommand{\Lya}{Ly$\alpha$}
\newcommand{\Lyb}{Ly$\beta$}
\newcommand{\MgII}{\hbox{Mg {\sc ii}}}
\newcommand{\CII}{\hbox{C {\sc ii}}}
\newcommand{\HI}{\hbox{H {\sc i}}}
\newcommand{\HII}{\hbox{H {\sc ii}}}
\newcommand{\NV}{\hbox{N {\sc v}}}
\newcommand{\OVI}{\hbox{O {\sc vi}}}

\newcommand{\NVb}{\hbox{$\rm NV_b$}}

\newcommand{\CIVb}{\hbox{$\rm CIV_b$}}
\newcommand{\sla}{\hbox{\small \textlambda}}
\newcommand{\lala}{\hbox{\small \textlambda\textlambda}}

\newcommand{\XR}{\hbox{X-ray}}
\newcommand{\XRs}{\hbox{X-rays}}
\newcommand{\PL}{\hbox{power-law}}

\newcommand{\asca}{{\emph{ASCA}}}
\newcommand{\chandra}{\emph{Chandra}}

\newcommand{\rosat}{{\emph{ROSAT}}}

\newcommand{\xmm}{\hbox{\emph{XMM-Newton}}}
\newcommand{\nustar}{\hbox{\emph{NuSTAR}}}

\newcommand{\vmin}{$v_{\rm min}$}
\newcommand{\vmax}{$v_{\rm max}$}
\newcommand{\vmean}{$v_{\rm mean}$}

\newcommand{\hst}{{\emph{HST}}}

\newcommand{\SB}{\hbox{0.5--2~keV}}
\newcommand{\HB}{\hbox{2--8~keV}}
\newcommand{\FB}{\hbox{0.5--8~keV}}

\newcommand{\Cstat}{\hbox{$C$-statistic}}
\newcommand{\xspec}{{\sc xspec}}

\newcommand{\cloudy}{{\sc cloudy}}

\newcommand{\appropto}{\mathrel{\vcenter{\offinterlineskip\halign{\hfil$##$\cr \propto\cr\noalign{\kern2pt}\sim\cr\noalign{\kern-2pt}}}}}
\title[{The X-rays wind connection in \pg}]
  {THE X-RAYS WIND CONNECTION IN PG~2112+059}
\author[{Saez et al.}]{
  C.~Saez $^{1}$,  
  W. N. Brandt$^{2}$, 
  F. E. Bauer$^{3,4,5}$, 
  G. Chartas$^{6}$,  
  T. Misawa$^{7}$, 
  F. Hamann$^{8}$,
  \newauthor
   and S. C. Gallagher$^9$  \\
 $^{1}$Departamento de Astronom\'ia, Universidad de Chile, Casilla 36-D, Santiago, Chile \\
$^{2}$  Department of Astronomy and Astrophysics, 525 Davey Lab, The Pennsylvania State University, University Park, PA 16802, USA \\
$^{3}$ Instituto de Astrof{\'{\i}}sica and Centro de Astroingenier{\'{\i}}a, Facultad de F{\'{i}}sica, Pontificia Universidad Cat{\'{o}}lica de Chile, Casilla 306, \\Santiago 22, Chile \\
$^{4}$ Millennium Institute of Astrophysics (MAS), Nuncio Monse{\~{n}}or S{\'{o}}tero Sanz 100, Providencia, Santiago, Chile \\
$^{5}$ Space Science Institute, 4750 Walnut Street, Suite 205, Boulder, Colorado 80301 \url{https://orcid.org/0000-0002-8686-8737} \\
$^{6}$ Department of Physics and Astronomy, College of Charleston, Charleston, SC 29424, USA \\
$^{7}$ School of General Education, Shinshu University, 3-1-1 Asahi, Matsumoto, Nagano 390-8621, Japan \\
$^{8}$ Department of Physics \& Astronomy, University of California, Riverside, CA 92507, USA \\
$^{9}$ Department of Physics and Astronomy and Institute for Earth and Space Exploration, The University of Western Ontario, London, \\ ON, N6A 3K7, Canada
 }
 \date{}

\pagerange{\pageref{firstpage}--\pageref{lastpage}} \pubyear{2020}

\def\LaTeX{L\kern-.36em\raise.3ex\hbox{a}\kern-.15em
    T\kern-.1667em\lower.7ex\hbox{E}\kern-.125emX}

\begin{document}

\label{firstpage}

\maketitle

\begin{abstract}

We study the connection between the X-ray and UV properties of the broad absorption line (BAL) wind in the highly \XR\ variable quasar \pg\ by comparing \chandra-ACIS data with contemporaneous UV \hst/STIS spectra in three different epochs. 
We observe a correlation whereby an increase in the  \hbox{equivalent-widths} (EWs) of the BALs is accompanied by a redder UV spectrum. The growth in the BALs EWs is also accompanied by a significant  dimming in soft X-ray emission ($\lesssim 2 \,\text{keV}$), consistent with increased absorption. 
Variations in the hard X-ray emission (\mbox{$\gtrsim 2\, \text{keV}$)} are only accompanied by minor spectral variations of the \hbox{UV-BALs} and do not show significant changes in the EW of BALs. These trends suggest a wind-shield scenario where  the outflow inclination with respect to the line of sight is decreasing and/or the wind mass is increasing. These changes elevate the covering fraction and/or column densities of the BALs and are likely accompanied by a nearly contemporaneous increase in the column density of the shield.

\end{abstract}

\begin{keywords}
techniques: spectroscopic --- X-rays: galaxies --- galaxies: active --- quasars: absorption lines ---  galaxies: individual: \pg.
\end{keywords}

\section{Introduction}
It is well established that, with exception of dwarf galaxies, every galaxy should have a supermassive black hole (SMBH) with black hole masses ($M_{\rm BH}$) of $10^6 M_\odot \lesssim M_{\rm BH} \lesssim 10^{10} M_\odot$ in their centers \citep[e.g.,][]{2018MNRAS.478.2576M,2013ARA&A..51..511K}. 
Active Galactic Nuclei (AGN), which in their brightest states become quasars\footnote{A quasar is defined as a ``bright AGN" with $M_B<-23$ or $L_{\rm B} \gtrsim 10^{44}\, \lumin$;  \citep{1983ApJ...269..352S} }, in brief \citep[of the order of $10^7$~years;][]{1993MNRAS.263..168H} and powerful duty cycles are thought to regulate SMBH growth \citep{1982MNRAS.200..115S}. AGN should also  transport energy from the SMBH to their surrounding host galaxy \citep[e.g,][]{2012ARA&A..50..455F} through various feedback mechanisms, including jets and quasar winds (or accretion disk winds). These two mechanisms are key ingredients to explain the regulation of galaxy evolution, and the origin of known observational relationships between the SMBH masses and bulge properties in nearby galaxies \citep[e.g.,][]{2006ApJS..163....1H, 2008MNRAS.391..481S, 2013ARA&A..51..511K}.

Evidence of winds is observed in the \hbox{ultra-violet} (UV) \RF\ spectra in a fraction of quasars \citep[$\sim 20\%$, e.g.,][]{2003AJ....125.1784H,2009ApJ...692..758G} through ionized broad absorption lines (BALs). These spectral features with broadening above $2000$~\kms\ \citep{1991ApJ...373...23W} appear as absorption with blueshifted velocity offsets of $2,000$ to $30,000$~\kms\ from the line \RF.
Depending on the ionization state of the wind, BAL quasars are divided into at least two categories \citep[e.g.][]{2002ApJS..141..267H}: high-ionization BAL quasars (HiBALs) and low-ionization BAL quasars (LoBALs). HiBALs show absorption features from \OVI, \NV, and \CIV\ lines. LoBALs are characterized to have high-ionization absorption plus absorption from \AlIII, \CII\  and/or \MgII\ lines. 

    Observations and theory suggest that quasar BAL features are an orientation effect \citep[e.g.,][]{1995ApJ...451..498M,2010A&A...516A..89R, 2014ApJ...791...88F}, and thus these outflows should be intrinsic to every quasar. Models predict that BALs originate from the accretion disk in funnel shaped structures at distances of $10^2-10^3 R_S$ (where $R_S=2GM_{\rm BH}/c^2$ is the Schwarzschild radius) from their central SMBH  \citep[e.g.,][]{1995ApJ...451..498M,2000ApJ...543..686P}. In contrast,  larger distances of $10^3-10^7R_S$ are inferred from the BAL observational signatures \citep[e.g.,][]{2001ApJ...548..609D,2011MNRAS.411.2653H,2013MNRAS.436.3286A}. Similar distance estimates are found for narrow absorption lines (NAL)  and mini-BALs quasars  \citep[e.g.,][]{2016ApJ...825...25M,2019ApJ...876..105X}.  Therefore, UV outflows could be observed far from their origin or a revision on their models is needed.  BAL quasars also tend to show particularly distinctive weak \XR\ emission \citep[e.g.,][]{2009ApJ...692..758G}, which has been attributed to absorption \citep[e.g.,][]{2017MNRAS.464.4586P} and/or intrinsically weak emission \citep[e.g.,][]{2014ApJ...794...70L}. 
    
The growing evidence from BAL quasar spectral features from UV to \XRs\ has motivated many models. These models postulate that outflows are launched in funnel shaped structures driven by radiation and/or magnetic forces from the accretion disk \citep[e.g.,][]{1995ApJ...451..498M,2000ApJ...543..686P,2010ApJ...723L.228F}. In view of these models, BALs correspond to lines of sight that are intercepting these structures. Additionally, the weak X-ray emission could be a product of absorption from the inner parts $\lesssim 100\,R_S$ of the medium that is not accelerated enough to be expelled to the inter galactic medium (IGM; failed wind). There could be also a highly ionized part of the wind accelerated to relativistic speeds $\gtrsim 0.1c$ (with $c$ the speed of light)  and showing spectral signatures in the \XR\ through highly ionized $\FeXXV$ and $\FeXXVI$ BAL lines \citep[e.g.,][]{2002ApJ...579..169C,2003ApJ...595...85C, 2011ApJ...737...91S,2018MNRAS.476..943H}.

The HiBAL quasar \pg\  with $z=0.459$ \citep[][]{2016AJ....152...25M}  and $M_V=-27.3$ is one of the most luminous PG quasars. It has been observed in \XRs\ from 1991-2007 by various \XR\ missions, including \rosat, \asca, \xmm, and \chandra. In these observations, \pg\ has shown typical BAL \XR\ weakness and order of magnitude \XR\ flux changes over periods of time as short as 6 months \citep[][]{2012ApJ...759...42S}. The physical model that leads to the observed X-ray emission is not well understood. It could be a \hbox{reflection-type} model \citep[e.g.,][]{2010A&A...512A..75S} or a complex absorption model \citep[e.g.,][]{2004ApJ...603..425G, 2012ApJ...759...42S}.
In this work,  our main goal is to analyze the connection between the strong X-ray variability of \pg\ and the physical properties that the associated BAL winds present in the UV. Through this kind of joint analysis, we are aiming to better understand the mechanisms that create these winds. 
Throughout this paper, unless stated otherwise, we use cgs units, errors are quoted at the 1$\sigma$ level, and we adopt a flat \hbox{$\Lambda$-dominated} universe with $H_0=70~\kms\;{\rm Mpc^{-1}}$, $\Omega_\Lambda = 0.7$, and $\Omega_M = 0.3$. 

\begin{figure}
   \includegraphics[width=8.4cm]{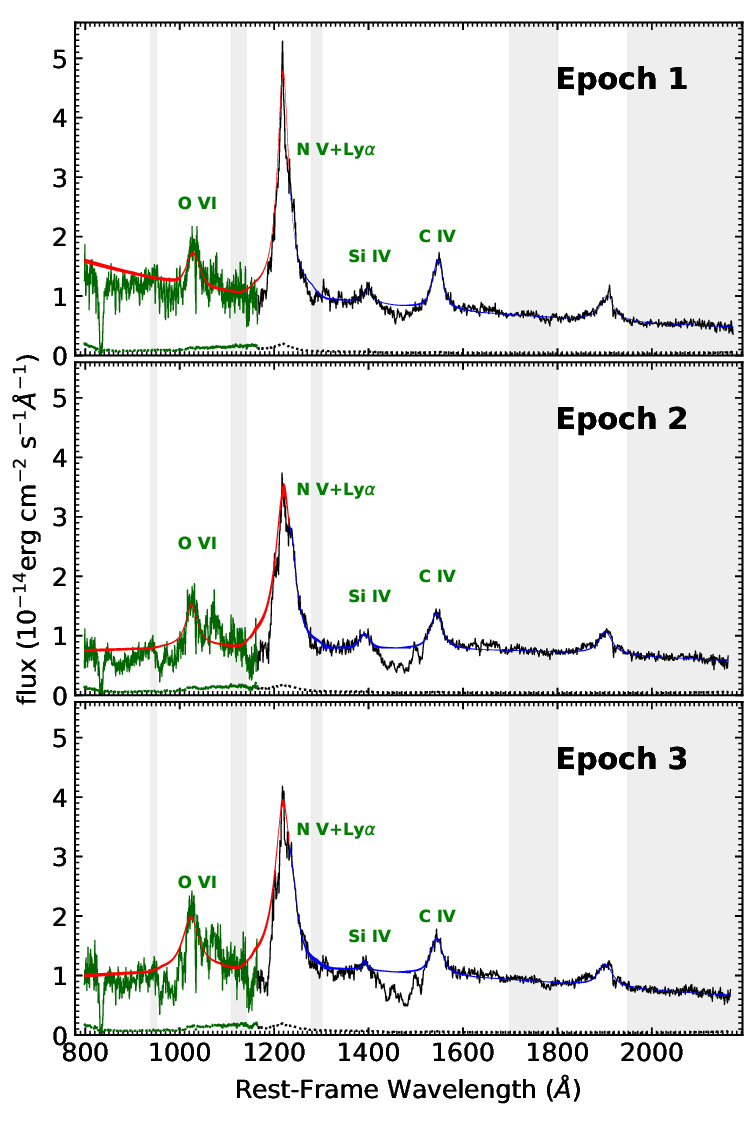}
        \centering
       \caption{\hst\ UV spectra of PG 2112+059 illustrating the presence of \OVI, \NV, and \CIV\ BALs. The upper, middle and lower panel correspond to  epochs 1, 2 and 3, respectively. The green and black curves indicate the G140L and G230L grating spectra, while, the red and blue curves are the fitted spectral model in the \hbox{extreme-UV} and \hbox{far-UV} (EUV and FUV) wavelength bands (see \S\ref{S:UVfit}) with thickness given by the 1$\sigma$ confidence region and the dotted line is the spectral error.  The gray areas are the wavelength ranges where continua fits are performed.}
      \label{fig:fitf}
     \end{figure}

\section{METHODOLOGY} \label{met}

\subsection{Observations}
\begin{table*}
\begin{minipage}{175mm}
\begin{center}
\caption{Log of \chandra\ observations of \pg \label{tab:chao}}
\begin{tabular}{lccccc}
\hline\hline
 &  
 & 
{\sc exp. time} &  
 & 
{\sc rate} \\
{\sc obs. date} & 
{\sc obs. id}  & 
{(ks)} & 
{\sc counts}  & 
{(10$^{-3}$~s$^{-1}$)} & 
{Ref.$^{\rm a}$} \\
\hline

2002/09/01\,(Epoch\,1) & 3011 & 56.9 & $829_{-29}^{+30}$ & 14.6$\pm$0.5 & 1  \\ 
2014/12/20\,(Epoch\,2) & 17553 & 18.2 & $62_{-8}^{+9}$ & 3.4$\pm$0.5 & 2  \\ 
2015/08/29\,(Epoch\,3) & 17148  & 18.3 & $140_{-12}^{+13}$ & 7.6$\pm$0.7 & 2  \\

\hline
\end{tabular}
\end{center}

All observations utilize the ACIS-S3 detector. The counts and rates are from background-subtracted source photon counts in the \chandra\ full-band (\FB). 
The exposure times, photon counts, and rates  are obtained after screening the data. The errors on the  source counts were computed by propagating the asymmetric errors on the total and background counts using the approach of \cite{2004physics...6120B}. The total and background count errors were estimated from Tables 1 and 2 of \cite{1986ApJ...303..336G}. 

$^{\rm a}$ {\sc References:} (1)~\cite{2004ApJ...603..425G}; (2)~This work. 

\end{minipage}
\end{table*}

\begin{table*}
\begin{minipage}{175mm}
\begin{center}
\caption{Log of \hst\ STIS observations of \pg \label{tab:hsto}}
\begin{tabular}{cccccccc}
\hline \hline
 & 
 & 
{\sc exp. time} &  
 &  \\
{\sc obs. date} & 
{\sc prop. id}  & 
{(s)} & 
{\sc instrument} & 
grating  & 
{\sc wave-range} & 
{\sc r$^{\rm a}$} & 
{Ref.$^{\rm b}$} \\
\hline
2002/09/01\,(Epoch\,1) & 9277 & 1100 & FUV-MAMA & G140L & 1150$-$1730 & 1223 & 1  \\
2002/09/01\,(Epoch\,1) & 9277 & 900 & NUV-MAMA & G230L & 1570$-$3180 & 767 & 1  \\ 
2014/12/18\,(Epoch\,2) & 13948 & 1095 & FUV-MAMA & G140L & 1150$-$1730 & 1222 & 2  \\
2014/12/18\,(Epoch\,2) & 13948 & 920 & NUV-MAMA & G230L & 1570$-$3180 & 767 & 2  \\ 
2015/09/11\,(Epoch\,3) & 13948 & 1095 & FUV-MAMA & G140L & 1150$-$1730 & 1222 & 2  \\
2015/09/11\,(Epoch\,3) & 13948 & 920 & NUV-MAMA & G230L & 1570$-$3180 & 767 & 2  \\
\hline
\end{tabular}
\end{center}
The STIS observations utilize the following gratings: G140L for the FUV-MAMA and G230L for the NUV-MAMA configurations.\\
$^{\rm a}$ The spectral resolution ($R=\lambda/\Delta \lambda$) is calculated at the central wavelength of each configuration, i.e. at 1425~\AA\ and 2376~\AA\ for the FUV-MAMA and NUV-MAMA configurations, respectively.\\
$^{\rm b}$ {\sc References:} (1)~\cite{2004ApJ...603..425G}; (2)~This work. 

\end{minipage}
\end{table*}

In this work, we analyze two $\sim20$~ks \chandra\ observations each with a contemporaneous \hst\ STIS spectrum, performed in 2014--2015 and separated by approximately nine months. These new observations are compared with a nearly simultaneous archival \chandra-\hst\ observation from 2002 \citep[analyzed in][]{2004ApJ...603..425G}.\footnote{Details about the motivation of these observations can be found at \cite{2016AN....337..541S}.} Each \chandra\ observation was performed within $\sim$10~days of an \hst\ observation. During the   \chandra-\hst\ time gap although there could be \XR\ spectral variability, this should not be important when compared to long term ($\gtrsim 6$~months) spectral variability. 
Based on the existing constraints for \pg\ we expect that any significant \XR\ spectral variability (relevant for this study) should be on time scales $\gtrsim1$~month. For instance, in the period between May 3 and November 5 of 2007 \pg\ has 4 \xmm\ observations each with at least 500 \hbox{0.3--10~keV} counts \citep[see][for details]{2012ApJ...759...42S}. During this period, no significant variability was detected (at a 99\% level), such that any potential flux variations were less than 20\%. 
Additionally, on time scales $\lesssim1$~month, it is expected that fractional changes in the normalized continuum flux removed by absorption should be $\lesssim 10$\% in the UV BALs \citep[e.g.,][]{2013MNRAS.429.1872C}. On time scales longer than a month the fraction of the normalized flux removed by absorption could change more dramatically in BALs. 
As we do not expect strong variability either in the \XR\ flux or in the UV BALs absorption profiles in scales $\lesssim 1$~month, hereafter, we will refer to each contemporaneous \chandra-\hst\ observation \hbox{time-ordered} epoch as 1, 2 or 3.
Details about the dates and most important characteristics of these three sets of \chandra-\hst\ spectra can be found in Table~\ref{tab:chao} for the \chandra\ observations and in Table~\ref{tab:hsto} for the \hst\ observations.

The \chandra\ observations were reduced  using the standard software CIAO version 4.12 provided by the \chandra\ X-ray Center (CXC). For each epoch, we reprocessed the datasets through the {\sc chandra\_repro} script to obtain the latest calibration.  
Source and background spectra and associated products were extracted using the CIAO script {\sc specextract}  from a circular region with an aperture radius of 4\arcsec\ and an annular source-free region with an inner radius of 6\arcsec\ and an outer radius of 24\arcsec, respectively. 
The numbers of background-subtracted counts in the source regions at energies of \FB\ and some details about the \chandra\ observations are presented in Table~\ref{tab:chao}.  

The \hst\ spectra were reduced with the standard STIS pipeline, and the resulting spectra are expected to have a flux calibration at the $\sim5$\% level of precision \citep{2011stis.book.....B}.
The \hst\ STIS spectra span a  wavelength range of 1570--3180~\AA\ which in the rest frame of \pg\ sample several important BAL features such as \OVI, Ly$\alpha$, \NV, \SiIV, and \CIV.
In each \hst\ observation epoch, the exposure was $\sim1100$~s and $\sim900$~s for the G140L and G230L gratings, respectively, (see Table~\ref{tab:hsto}) both with a $52\arcsec \times  0\farcs2$ slit. 
The spectra have dispersions on average of $\approx0.58$~\AA\ and $\approx1.54$~\AA\ per pixel for the G140L and G230L gratings, and resolving powers of $R\approx1220$ and $R\approx770$ at the grating central wavelengths of 1425~\AA\ and 2376~\AA, respectively.
The G140L and G230L grating spectra overlap in the range 1570--1730~\AA. 
In this wavelength overlap region, as a product of the calibration, each grating spectra show decreasing signal-to-noise ratios (S/N) per pixel at their respective endpoints. Therefore, for our analysis,  we select a limiting wavelength of 1700~\AA\ ($\approx 1170$~\AA\ in the \RF\ of \pg) as the end of the G140L spectrum and beginning of the G230L spectrum. This wavelength is approximately where both gratings reach the same S/N in wavelength bins of comparable sizes. {In each observation, the S/N in the continuum is on average $\approx10$ (19) per pixel with a standard deviation of $\approx3$ (5) for the G140L (G230L) grating.} The full reduced and dereddened \hst\ spectrum of \pg\ at each epoch is presented in Figure~\ref{fig:fitf}. The dereddening has been obtained by assuming a Galactic extinction of $E(B-V)=0.0904$ \citep[from][]{1998ApJ...500..525S}, equivalent to $N_{H}\approx6 \times10^{20}\,$\cmsq\ \citep[e.g.,][]{2009MNRAS.400.2050G}, where $N_{\rm H}$ is the total (neutral an ionized) hydrogen column density. The reddening model used here and hereafter is that of \cite{1992ApJ...395..130P}. 
\begin{table}
\begin{center}
\caption{\chandra\ \XR\  spectral fits statistics \label{tab:fits}}
\begin{tabular}{lccc}
\hline\hline
 &  
\multicolumn{2}{c}{\Cstat/dof} & 
\\
{\sc epoch} & 
{APL}  & 
{WADR} & 
{$p$-values} \\
\hline
1 & 504.0/509 & 459.7/508 & 0.000 \\
2 & 235.8/509 & 231.1/508 & 0.007 \\
3 & 374.9/509 & 365.7/508 & 0.002 \\


\hline
\end{tabular}
\end{center}
%

 

\end{table}

\begin{table}
\begin{center}
\caption{\chandra\ \XR\  spectral fits parameters of the WAPLR model \label{tab:fpar}}
\begin{tabular}{ccccc}
\hline\hline
{\sc epoch} & 
{\nh}  & 
{log~$\xi$} & 
{norm$_{\rm PL}$} &
{norm$_{\rm R}$} \\
{(1)} &
{(2)} &
{(3)} &
{(4)} & 
{(5)} \\

\hline
1 &   12$\pm$  2 & 2.14$\pm$0.03 &  6.4$\pm$1.0 & 14.6$\pm$5.5 \\ 
2 &   64$\pm$ 42 & 3.02$\pm$1.49 &  1.4$\pm$0.3 & 13.9$\pm$7.7 \\ 
3 &   16$\pm$  2 & 1.95$\pm$0.03 &  7.5$\pm$1.4 &  1.6$\pm$5.3 \\ 

\hline
\end{tabular}
\end{center}

Col. (1): Observation Epoch. Col. (2): Column density in units of $10^{22}$~\cmsq. Col. (3): Logarithm of the ionization parameter $\xi=L/n \,r^2$, as defined in \xspec. \hbox{Col. (4--5)}: Normalizations of the continuum \PL\ and the reflection model ({\sc pexmon}). Both normalizations are in units of $10^{-5}\,{\rm counts\: \,cm^{-2}\,s^{-1}\,keV^{-1}}$.

\end{table}

\begin{figure}
   \includegraphics[width=8.4cm]{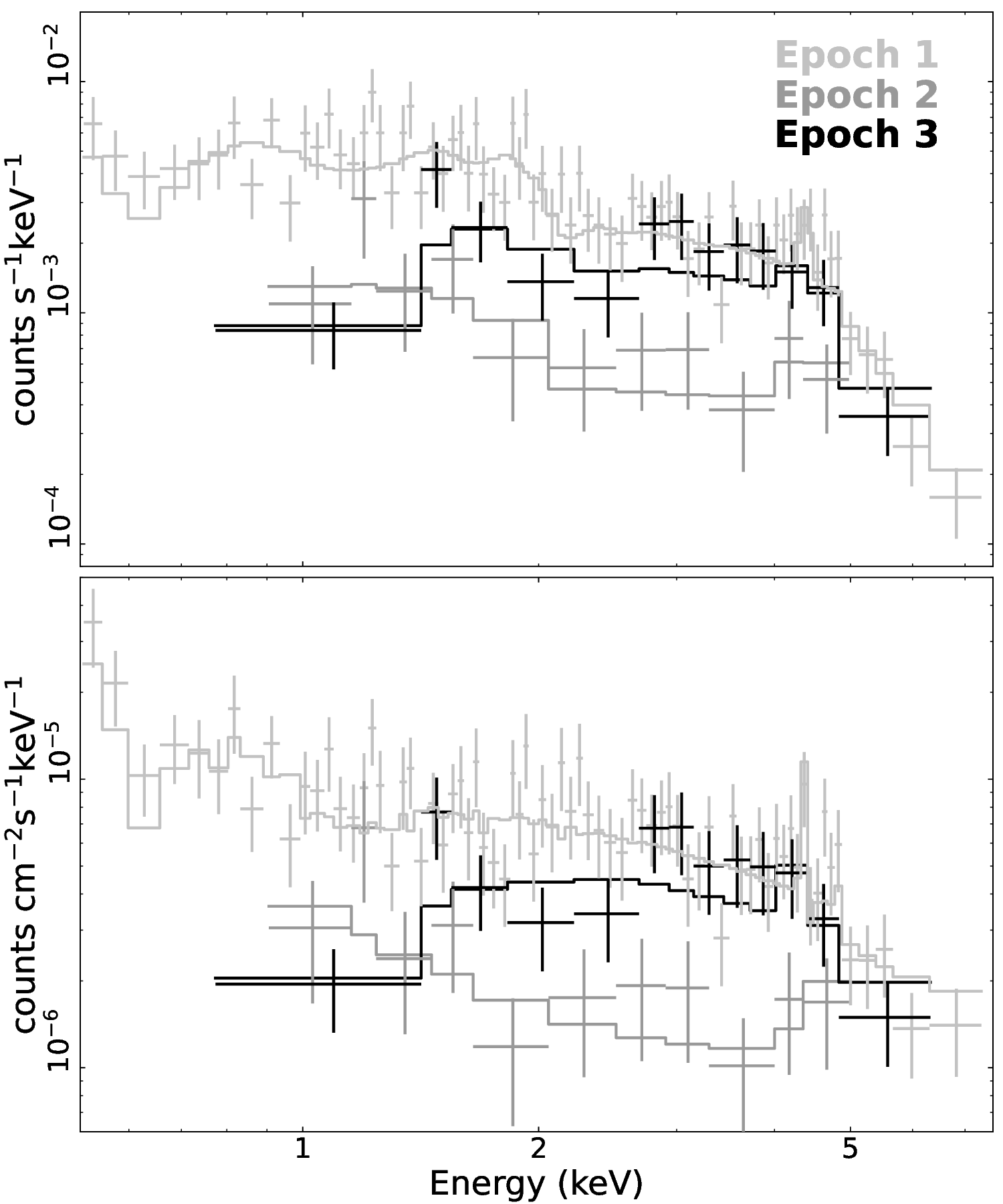}
        \centering
       \caption{Folded (upper panel) and unfolded (lower panel) \chandra\ spectra of \pg. The solid
        histograms indicate the best-fitting WAPLR model to each observation. The light gray, gray, and black colors are associated with epochs 1, 2 and 3, respectively. In each observation a grouping with a minimum 10 counts per bin has been used.}
     \label{fig:xrsp}
     \end{figure}

\subsection{X-ray analysis} \label{S:XRan}

Given that both of our new \chandra\ observations of \pg\ have very modest count statistics, we performed fits in the \FB\ band using the \Cstat\ \citep{1979ApJ...228..939C} in \xspec\ \citep{1996ASPC..101...17A}. Additionally, in all the \XR\ models used here Galactic absorption with total hydrogen column density of $\nh =6.0 \times 10^{20}$~\cmsq\ is assumed \citep{2005A&A...440..775K}. We use unbinned data and thus the \Cstat\ may not be appropriate
	to utilize when performing background subtraction. For this purpose, we fit the background
	spectra with a flat response using the {\sc cplinear} model \citep{2010ApJ...714.1582B}. The background model is scaled and subtracted when we fit the source spectra.
Our new \XR\ observations were designed to provide flux constraints, and with this goal we use a model that is simple and at the same time provides good fits to our observations.

To  fit the \chandra\ spectra, we select a model used in \cite{2010A&A...512A..75S} consisting of a continuum power law and an ionized reflection component, these both pass through a warm absorber. Hereafter we will refer to this model as the \hbox{warm-absorbed} \PL\ and reflection model (WAPLR; \xspec\ model {\sc phabs*zxipcf*(pow+pexmon)}).  To avoid overfitting our spectra we fixed many of the parameters that describe the WAPLR model\footnote{For the warm absorber (\xspec\ model {\sc zxipcf}),  we fix the covering fraction to 1. For the continuum \PL\ model (\xspec\ model {\sc pow}), we fix the photon index to 1.9.  The value of $\Gamma=1.9$ is close to an average value of the photon index for quasars \citep[e.g.,][]{2000MNRAS.316..234R, 2008AJ....135.1505S,2017A&A...603A.128N}. For the reflection model (\xspec\ model {\sc pexmon}), we fix the incident \PL\ photon index to 1.9,  the cutoff energy to 150~keV, the scaling factor for reflection to $-1$ (no direct emission, only reflected component),  the inclination angle to $45^\circ$, and we assume solar abundances. }, leaving 4 degrees of freedom: the normalization of the continuum \PL\ (${\rm norm_{PL}}$), the normalization of the reflection component (${\rm norm_{R}}$), and  the column density (\nh) and the logarithm of ionization parameter (as defined by \xspec; $\log \xi$) of the warm absorber.  

The selection of the WAPLR model over a simpler model like the absorbed \PL\ model (APL;\footnote{The APL model consist on a \PL\ with an intrinsic neutral absorber component, and thus it has 3 degrees of freedom: the normalization and photon index of the \PL, and the hydrogen column density (\nh) of the redshifted absorber.} \xspec\ model {\sc phabs*zphabs*pow}) is based on evidence of spectral complexity that includes the presence of a reflection component  in previous \pg\ high S/N spectra   \citep[e.g.,][]{2010A&A...512A..75S, 2012ApJ...759...42S}. Additionally, the WAPLR model is a good representation of the data in past \xmm\ observations of \pg\ \citep{2007A&A...474..431S, 2010A&A...512A..75S}, and  provides significant improvements over the fits of an APL model over Epochs~\hbox{1--3} (see next paragraph).  In Figure~\ref{fig:xrsp} we present the \chandra\ spectra  with their respective WAPLR fits. As this figure shows, the WAPLR model reproduces  the significant Fe~K$\alpha$ emission feature at \RF\ energies $\sim 6.4~{\rm keV}$ (feature at $\sim 4.4$~keV in Figure~\ref{fig:xrsp}) found in Epoch~1 \citep{2004ApJ...603..425G}. 

In Epochs~\hbox{1--3} we check that the WAPLR model shows an improvement in the fits (lower values of \Cstat; see Table~\ref{tab:fits}) when compared with the APL model. 
We test these fit improvements by using a modified version of the likelihood ratio test (LRT) routine provided by the \xspec\ package.\footnote{\burl{https://heasarc.gsfc.nasa.gov/xanadu/xspec/manual/node124.html}.}  Using this routine we generate 1000 Monte Carlo simulated spectra from the fits of the APL  model. These simulated spectra are fitted with both the APL and WAPLR  models to generate a table of $\Delta C$, where $\Delta C$ is the difference between values of the \Cstat\ obtained with the APL and the WAPLR model. The $p$-values correspond to the fraction of fits where $\Delta C > \Delta C_{\rm obs}$, where $\Delta C_{\rm obs}$ is the observed value of $\Delta C$. 
From the simulated data, in Epochs~1 and 3 we check that the WAPLR provides a significant improvement over the fit with the APL model ($p  \lesssim 0.002$)   (see Table~\ref{tab:fits}). We also find in Epoch~2 a significant improvement on the fits (with $p \approx 0.007$), albeit, this observation might not have enough counts to be reliable in differentiating between model fits. The best-fitted parameters and fluxes obtained from the WAPLR model are presented in Tables~\ref{tab:fpar} and \ref{tab:flxx}.  
Hereafter (if not stated otherwise), we assume that any \XR\ parameter estimated from spectral fits is obtained from WAPLR model using the \Cstat.

\begin{table*}
\begin{minipage}{175mm}
\begin{center}
\caption{X-ray fluxes and Luminosities of \pg  \label{tab:flxx}}
\begin{tabular}{cccccccccccc}
\hline\hline
{\sc epoch} &
{$f_{0.5-2}$} &
{$f_{2-8}$} &
{$f_{0.5-8}$} & 
{$f_{\rm 2keV}$} &
{$\Gamma_{\rm eff}$} &
{\textsc HR} &
{${\rm log}\,L_{\rm 2keV}$ } &
{$\lLa$}  \\
{(1)} &
{(2)} &
{(3)} &
{(4)} &
{(5)} &
{(6)} &
{(7)} &
{(8)} &
{(9)} 
 \\
\hline
1 &  3.2$\pm$0.2 &   16$\pm$  1 &   19$\pm$  1 &  5.5$\pm$0.4 & 0.85$\pm$0.08 & 0.21$\pm$0.04 & 25.47$\pm$0.03 & 44.07$\pm$0.03 \\ 
2 &  1.1$\pm$0.5 &    4$\pm$  2 &    5$\pm$  3 &  1.5$\pm$0.7 & 1.00$\pm$0.61 & 0.15$\pm$0.14 & 24.90$\pm$0.21 & 43.48$\pm$0.17 \\ 
3 &  1.2$\pm$0.2 &   13$\pm$  2 &   14$\pm$  2 &  2.5$\pm$0.3 & 0.29$\pm$0.14 & 0.43$\pm$0.08 & 25.12$\pm$0.06 & 43.96$\pm$0.06 \\ 

%
\hline
\end{tabular}
\end{center}

X-ray fluxes, luminosities and total hydrogen column densities are corrected for Galactic absorption assuming $\nh=6.0 \times 10^{20}$~\cmsq\  \citep{2005A&A...440..775K} and are obtained from the best-fitting parameters for a 
WAPLR model (\xspec\ model {\sc phabs*zxipcf*[pow+pexmon]}). Col. (1): Observation Epoch. Cols. (2--4): Observed fluxes in the  \SB, \HB, and \FB\  bands in units of $10^{-14}$~\flux. Col. (5): Observed flux densities at \RF\  2~keV in units of $10^{-32}$~\flux Hz$^{-1}$. Col. (6): X-ray effective photon index $\Gamma_{\rm eff}=-\alpha_x+1$ obtained from the ratio of of \HB\ to \SB\ flux assuming \PL\ spectra ($f_\nu \propto \nu^{\alpha_x}$). Col. (7): Hardness ratio,  corrected due to time-dependent loss of \chandra-{\emph ACIS} sensitivity (see \S\ref{S:resu} for details).
 Col. (8) Logarithm of the monochromatic luminosity at \RF\ 2 keV. Col. (9) Logarithm of the luminosity at \RF\ 2--10~keV. 
\end{minipage}
\end{table*}

\subsection{UV Continuum Fit and Normalization} \label{S:UVfit}

To describe the \CIV\ BAL, we fit the \hst/UV continuum in three relatively line-free (RLF) windows at \RF\ wavelengths in the far-UV band (FUV;  \hbox{$>1200$~\AA}) of \pg:  \hbox{1280--1300~\AA}, \hbox{1700--1800~\AA}, and \hbox{1950--2200~\AA}. These RLF windows were selected in a similar fashion as  \cite{2009ApJ...692..758G}, with a small difference in the first RLF window.  For this window, the lower wavelength limit is 1280\AA\ (instead 1250\AA) in order to avoid the broad Ly$\alpha$+\NV\ emission line. Additionally,  the upper limit is 1300\AA\ (instead 1350\AA) in order to avoid a possible BAL \SiIV\ region.
The continuum in the extreme-UV (EUV) band ($< 1200$~\AA)  is expected to have thermal signatures of the accretion disk \citep[i.e., big-blue-bump features; e.g.][]{2002ApJ...565..773T,1997ApJ...475..469Z} and to exhibit significant unaccounted attenuation mainly due to the Lyman forest\footnote{It is a product of intervening lines \Lya\ \sla1216, \Lyb\ \sla1026 and the depression of the continuum at wavelengths close to the Lyman break limit (at 912~\AA).   Individual intervening \HI\ absorption have been identified in high-resolution UV observations of quasars at $z\sim 0.3$  \citep[see e.g.,][]{2007ApJ...658..680L}. Additionally, in quasars at $z\gtrsim 0.3$ it is expected attenuation of the continuum $\sim 10\%$ at wavelengths nearby the Lyman break limit (approximately in the range [800--900]~\AA) \citep[e.g.,][]{1997ApJ...475..469Z}.}, i.e., the integration of  \HI\ absorption features 
 coming from intervening media. At the redshift of \pg\ we expect that these features show approximately in the wavelength range [800--1200]. Therefore, to study the \OVI\ and \NV\  BALs, we independently characterize the EUV band through two RLF windows: \hbox{940--950~\AA},  and \hbox{1110--1140~\AA}. 
The use of the \hbox{940--950~\AA} window is to constrain the fitted continua to be close to the spectra  at a wavelength $\sim945$~\AA. This wavelength, as seen from Figure~\ref{fig:fitf}, should be close to the blue end of the \OVI\ BAL region \citep[see, for example,][]{2017MNRAS.468.4539M}. Additionally, the \hbox{1110--1140~\AA} window is coincident with the blue end of the \NV\ BAL. The RLF windows used to fit the FUV (above 1200\AA) and EUV (below 1200\AA) bands are shown in Figure~\ref{fig:fitf} as gray areas.

The selected RLF windows described in the last paragraph are fitted using a least-squares sigma-clipping method, discarding data that deviate by greater than  3$\sigma$ from the model. In the FUV band,  to estimate the continua, we use a model composed of a \PL\
intrinsically reddened. There are only three parameters in this model: \PL\ normalization, the \PL\ spectral index, and the magnitude of intrinsic reddening E(B-V). Given that there is degeneracy between the UV emission continuum shape and E(B-V), we do not physically interpret the fitted values of the intrinsic reddening. Additionally, in the FUV band, we also performed fits using a unreddened (unabsorbed) \PL.
These fits were used to estimate the steepness of the spectra through the spectral index $\alpha_{FUV}$, the monochromatic fluxes at 2000~\AA, and to extrapolate monochromatic fluxes at 2500~\AA; all these parameters are listed in Table~\ref{tab:fluv}.  
For the EUV band, given the reduced wavelength range of the fitting regions, a simple, unreddened \PL\ were used.  The monochromatic fluxes at 950~\AA\ and  the spectral indexes $\alpha_{EUV}$ obtained with these fits are also found in Table~\ref{tab:fluv}.  Note that all spectral indexes presented in Table~\ref{tab:fluv} are in frequency units.\footnote{A power law continuum is  expressed in frequency units as $f_\nu\propto \nu^{\alpha}$ where $\alpha$ is the spectral index. A  \PL\ continuum in  frequency units $f_\nu\propto \nu^{\alpha}$ is a \PL\ continuum in wavelength units $f_\lambda \propto \lambda^{\alpha_\lambda}$ with $\alpha_\lambda=-(\alpha+2)$.}
The BAL features obtained by subtracting the fitted continuum still have second order deviations near their borders caused by strong broad emission lines (ELs). To mitigate this, the strongest emission lines (i.e., \CIV, \SiIV, Ly$\alpha$+\NV, and \OVI; ELs) were fitted using Voigt profiles, and the resulting fitted model spectra (continuum+ELs), which is assumed to be absorbed by the BALs, can be seen in Figure~\ref{fig:fitf}.   This figure highlights that there is no significant \SiIV\ BAL feature observed at any epoch.  

\begin{table*}
\begin{minipage}{175mm}
\begin{center}
\caption{UV properties of \pg  \label{tab:fluv}}
\begin{tabular}{cccccccccccc}
\hline\hline
{\sc epoch} &
{$f_{950}$} &
{$f_{2000}$} &

{$f_{2500}$} & 
{$AB_{2500}$} & 
{${\rm log}\,L_{2500}$ } &
{$\alpha_{EUV}$} &
{$\alpha_{FUV}$} &
{\aox} &
{$\Delta\aox$} \\
{(1)} &
{(2)} &
{(3)} &
{(4)} &
{(5)} &
{(6)} &
{(7)} &
{(8)}   &
{(9)} &
{(10)} \\
\hline
1 & 0.83$\pm$0.05 & 1.56$\pm$0.08 & $1.79\!\pm\!0.09$ & 15.77$\pm$0.06 & 30.98$\pm$0.02 & $-0.82\!\pm\!0.13$ & $-0.62\!\pm\!0.03$ &  $-2.12\!\pm\!0.02$ & $-0.48$ \\
2 & 0.51$\pm$0.04 & 1.86$\pm$0.09 & $2.39\!\pm\!0.12$ & 15.45$\pm$0.06 & 31.11$\pm$0.02 & $-2.31\!\pm\!0.16$ & $-1.15\!\pm\!0.03$ &  $-2.38\!\pm\!0.08$ & $-0.73$ \\
3 & 0.68$\pm$0.05 & 2.17$\pm$0.11 & $2.62\!\pm\!0.13$ & 15.35$\pm$0.05 & 31.15$\pm$0.02 & $-2.40\!\pm\!0.14$ & $-0.85\!\pm\!0.03$ &  $-2.31\!\pm\!0.02$ & $-0.66$ \\

\hline
\end{tabular}
\end{center}

Details about this table are provided in \S\ref{S:UVfit}. \hbox{Rest-frame} Monochromatic fluxes (Cols. 2--4) are obtained from continuum fits of dereddened spectra using the Galactic extinction given by $E(B-V)=0.0904$ \citep[from][]{1998ApJ...500..525S} and  are in units of  $10^{-26}$~\flux Hz$^{-1}$. Errors in monochromatic fluxes are obtained by adding in quadrature the 1$\sigma$ errors associated with continuum fits with the expected 5\% errors associated 
with flux calibration \citep{2011stis.book.....B}.
Col. (1): Observation Epoch. Col. (2):  \hbox{Rest-frame} monochromatic flux at 950~\AA. Col. (3): \hbox{Rest-frame}  monochromatic flux at 2000~\AA. Col. (4):  \hbox{Rest-frame}  monochromatic flux at 2500~\AA.   Col. (5): \hbox{Rest-frame}  Monochromatic AB magnitude at 2500~\AA. Col. (6): Logarithm of monochromatic luminosity  at rest-frame wavelength 2500~\AA. 
Col. (7): \hbox{Power-law} UV index at the extreme-UV (EUV; $<1200$~\AA). Col. (8): \hbox{Power-law} UV index at the far-UV (FUV; $>1200$~\AA). Col. (9): Galactic absorption corrected optical-to-X-ray \PL\ slope $\aox=0.384\,{\rm log}(f_{\rm 2keV}/f_{2500\Angst})$. No correction for intrinsic reddening or absorption has been made. Col. (10): The difference between the measured \aox\ value and the predicted  \aox\ for radio quiet quasars, based on the \cite{2007ApJ...665.1004J} relation: $\aox =-0.140\,{\rm log} L_{2500\Angst} + 2.705$.

\end{minipage}
\end{table*}

\begin{figure}
   \includegraphics[width=8.4cm]{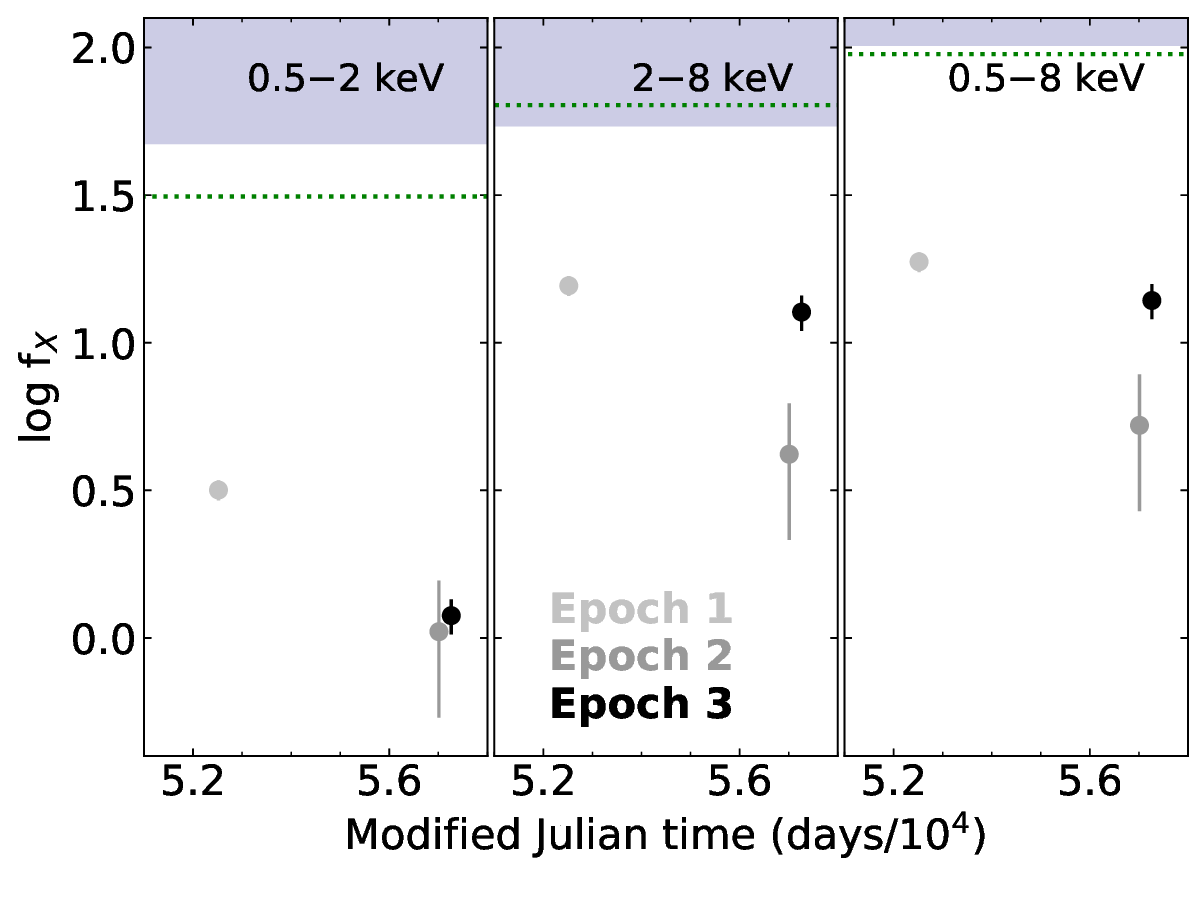}
        \centering
       \caption{Logarithm of the observed flux in the \chandra\ soft (\SB), hard (\HB), and full (\FB) bands in units of $10^{-14}$~\flux\ versus modified Julian date for each \chandra\ observation of PG 2112+059. The shaded area corresponds to the expected range of X-ray fluxes obtained assuming a typical $\aox=0.384\times {\rm log}[L_{\rm 2keV}/L_{2500}]$ \citep[see Table 5 of][]{2006AJ....131.2826S} and a \PL\ spectrum with $\Gamma=1.9$. The horizontal dotted line corresponds to the best-fit brightest flux ever obtained (in 1999 with ASCA, MJD$=51481$). The light gray, gray and black colors are associated with epochs 1, 2 and 3, respectively.}
     \label{fig:flxx}
     \end{figure}

\section{Results} \label{S:resu}

 In this section, we  test the significance of the variability of different spectral (UV and \XR) parameters between two epochs. In order to do this, we calculate the $\chi^2$ statistic assuming that the data points are the estimated parameter  (with their respective 1$\sigma$ errors), and the model is a constant obtained from the best fit. The $\chi^2$ value provides a statistical test of the null hypothesis that the parameter value of each epoch is equal to its best-fitted value. This model has one parameter to fit  two data points, thus the $\chi^2$ value follows a $\chi^2$-distribution with one degree of freedom. From hereafter, we refer to a significant (marginal) change of a parameter between two epochs, when the null hypothesis probability is less than 0.01 (between 0.05 and 0.01).
The soft \XR\ fluxes (i.e., at energies up to 2~keV) in Epoch~1 are significantly larger than those in the other epochs (see Table~\ref{tab:flxx}). Additionally, when Epochs~2 and 3 are compared, although soft \XR\ fluxes  do not show significant variation, the hard flux presents a marked increase (see Figures~\ref{fig:xrsp}--\ref{fig:flxx} and Table~\ref{tab:flxx}). We also find hardening in the \XR\ spectra when Epoch~1 is compared with Epoch~3 with a significant increase in the hardness-ratio (hereafter HR\footnote{Hardness ratio defined as HR=(Hc-Sc)/(Hc+Sc); where Hc and Sc are the source counts in the hard band (\HB) and soft band (\SB) respectively. In the values of HR presented in Table~\ref{tab:flxx},  the counts have been corrected due to time-dependent loss of \chandra-{\emph ACIS} sensitivity by using the effective area of Epoch~3 as reference. }) as seen in Table~\ref{tab:flxx}. 
As compared to the X-rays, the UV continua show only minor variations (see Figure~\ref{fig:fitf} and Table~\ref{tab:fluv}), and thus, the multiwavelength changes (measured by \aox) between Epoch~1 and the other epochs, are mainly the product of the significant variation in the soft X-ray flux. 
Additionally,  a significantly redder UV spectrum is observed when Epoch~1 is compared with Epochs~2-3  (see \aeuv\ and \afuv\ in Table~\ref{tab:fluv}).
\begin{figure}
	\includegraphics[width=8.4cm]{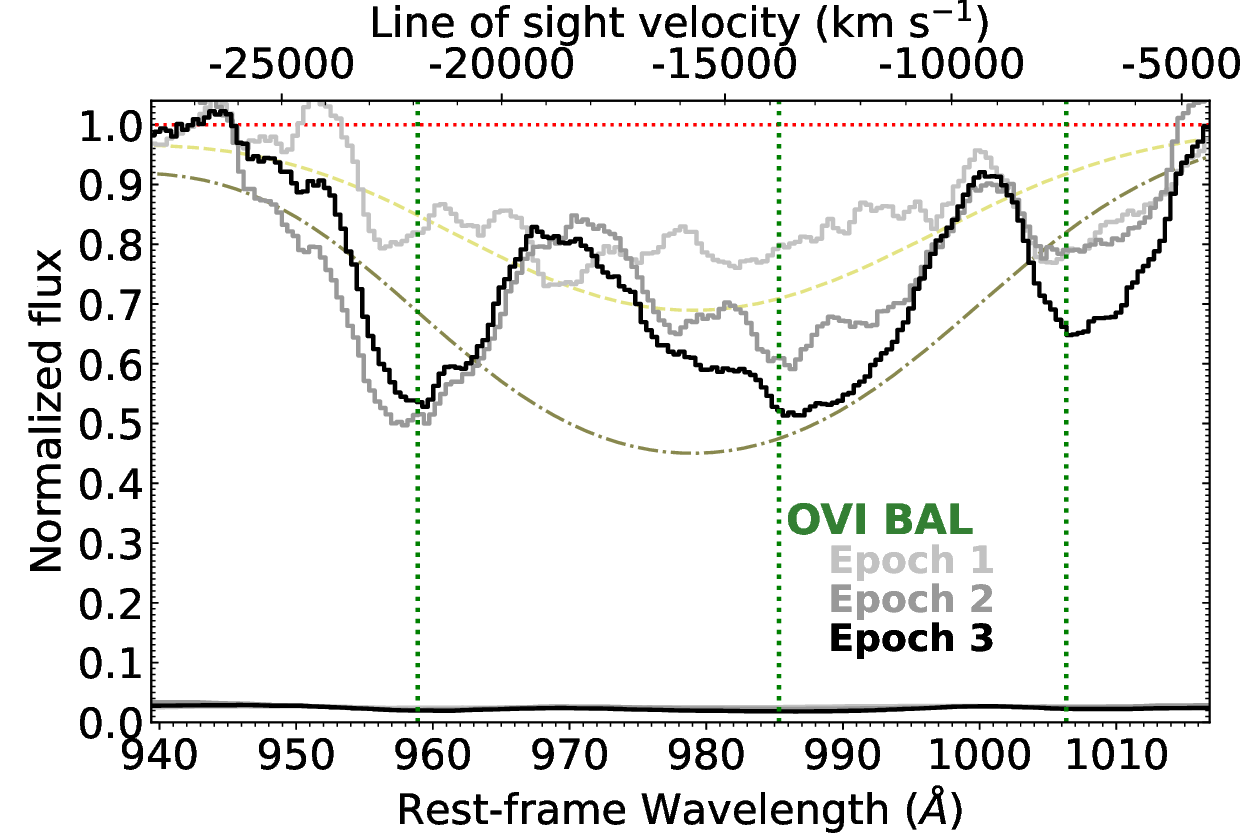}
	\centering
	\caption{Normalized flux in the \OVI\ BAL region. The light gray, gray and black colors histograms are the STIS spectra observed in epochs 1, 2 and 3, respectively. The lower curve correspond to errors at the 68\% confidence level. The three vertical dotted lines indicate line of sight outflow velocities of $-21900$,$-13800$ and $-7500$~\kms\ around local minima respectively. The dashed and \hbox{dot-dashed} curves correspond  to blue-shifted (blue-shift velocity of 16000~\kms) partially covered ($C_f=0.8$) absorption  profiles   obtained from \cloudy\ runs with $\log U=-0.8$, turbulent velocity 6500~\kms\ and column densities $\lnh=20.1$ and $\lnh=20.5$ respectively. Figures~\ref{fig:ovin}--\ref{fig:civn} are obtained using boxcar smoothing over 13 pixels for the G140L spectra and 7 pixels for the G230L spectra.
	}
	\label{fig:ovin}
\end{figure}

\begin{figure}
	\includegraphics[width=8.4cm]{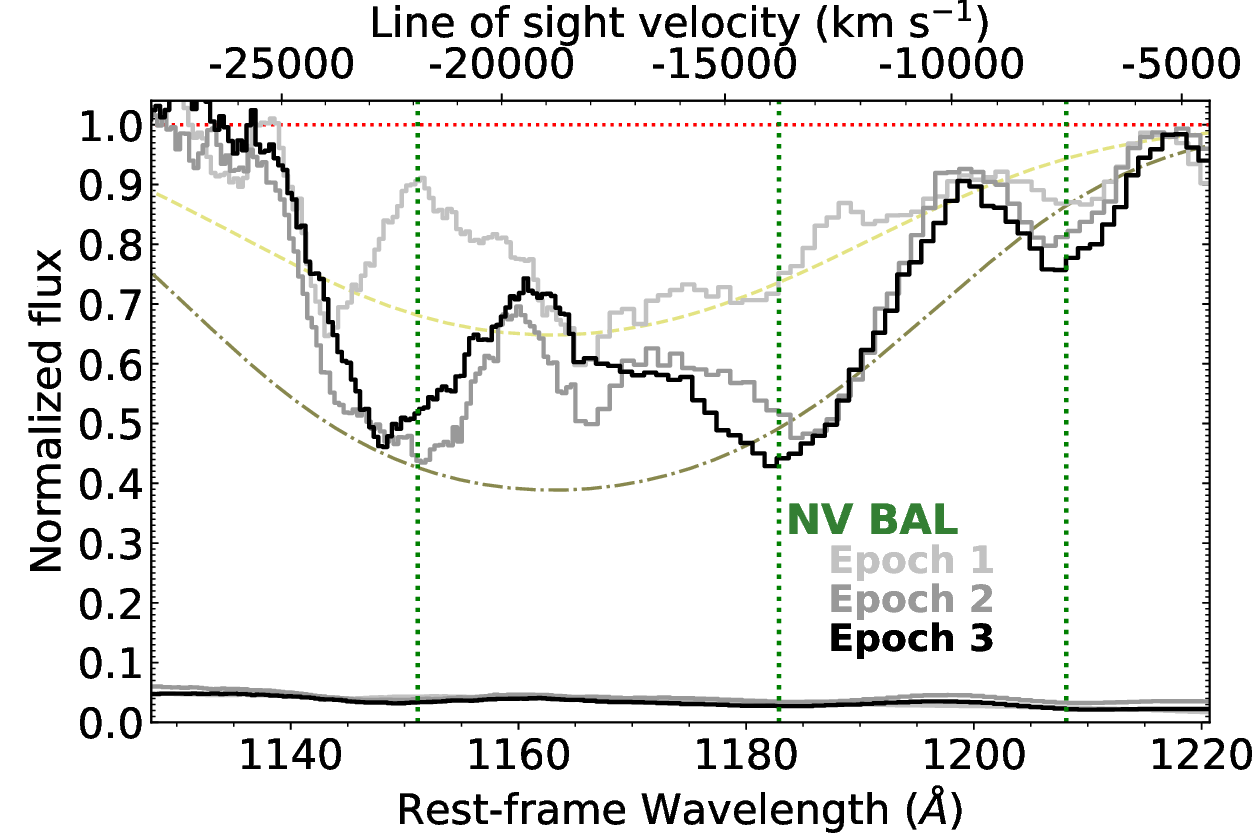}
	\centering
	\caption{Normalized flux in the  Ly$\alpha$-\NV\ BAL region (see legend of Figure \ref{fig:ovin} for more details).
	}
	\label{fig:lyan}
\end{figure}

\begin{figure}
	\includegraphics[width=8.4cm]{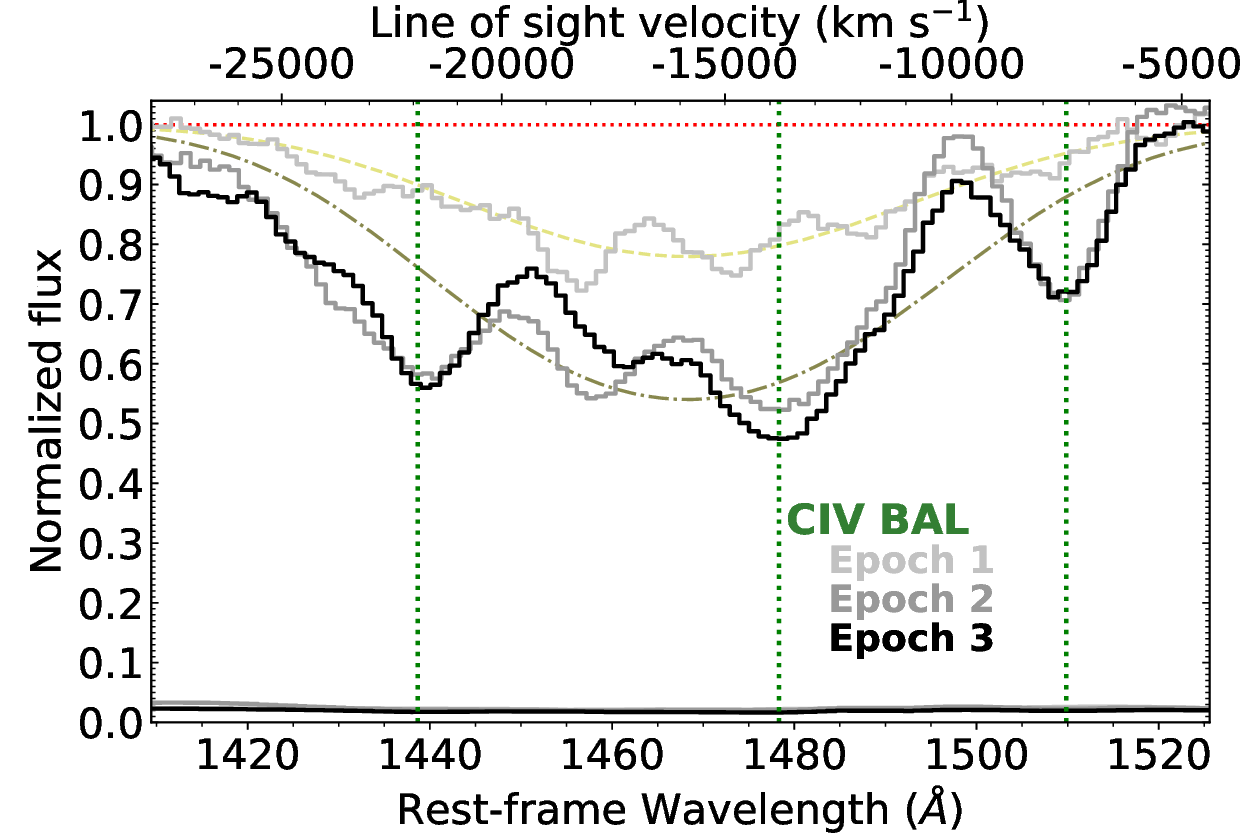}
	\centering
	\caption{Normalized flux in the \CIV\ BAL region  (see legend of Figure \ref{fig:ovin} for more details). 
	}
	\label{fig:civn}
\end{figure}

 The observed BALs in the \hst\ spectra are mainly due to the  \OVI, \NV, and \CIV\ line doublets, albeit there are other absorption lines as well that could have secondary contributions in the observed BALs (see \S\ref{S:clou} for details). Hereafter, unless otherwise stated, we assume that the zero velocity for each ion is the absorbing line laboratory wavelength for the blue component of the doublets given by \cite{1994A&AS..108..287V}. These wavelengths are 1031.9~\AA\ (\OVI), 1238.8~\AA\ (\NV), and 1548.2~\AA\ (\CIV).  
Using the fitted model spectra (continuum+ELs) described in \S\ref{S:UVfit} we obtain the normalized spectra around the \OVI, \NV, and \CIV\ BALs as shown in Figures~\ref{fig:ovin}--\ref{fig:civn} . 
In general, for all the BALs observed, Epochs~2 and 3 show more conspicuous absorption than Epoch~1. Additionally, the spectral differences are most pronounced when comparing either Epoch~2 or 3 with Epoch~1. 
 Epochs~2 and 3 show strong differences with Epoch~1 that are concentrated at $\approx960$~\AA\  and $\approx990$~\AA\ for the \OVI\ BAL, at $\approx1150$~\AA\ and $\approx1190$~\AA\  for the \NV\ BAL, and at $\approx1440$~\AA\ and $\approx1480$~\AA\ for the \CIV\ BAL. 
For every ion, when Epoch~2 or 3 are compared with Epoch~1, the zones of variability are more intense at velocities $\approx -22000\:\kms$ and $\approx -13000\:\kms$.
Epochs~2 and 3 show small differences between each other that seem to be more marked at $\approx990/1010$~\AA\ ($\approx-12000/-7000\:\kms$), $\approx1180$~\AA\ ($\approx-14000\:\kms$) and $\approx1460$~\AA\ ($\approx-17000\:\kms$) for the \OVI, \NV, and \CIV\ BALs, respectively.
The normalized spectra also show a reasonable level of consistency when observed  as a function of velocity. For example, the BALs in Epoch~1 appear as relatively featureless profiles in the range of velocities between $-25000$ to $-5000$~\kms. On the other hand, the absorption profiles of Epochs~2 and 3 seem to share similar distinguishable features centered at speeds $\sim -22000$, $\sim -14000$ and $\sim -8000$~\kms\ respectively (see vertical dotted lines in Figures~\ref{fig:ovin}--\ref{fig:civn}).
As a separate note, an \hst\ FOS observation performed in 1992 shares qualitative spectral similarities with Epochs~2 and 3. In this observation the UV spectral slope of \pg\ was {\bf redder}, and the EWs of the BALs were greater when  compared with Epoch~1 \citep[see Fig. 4 of][]{2004ApJ...603..425G}.
\begin{table*}
\begin{minipage}{175mm}
\begin{center}
\caption{Equivalent widths of UV BALs \label{tab:EqWi}}
\begin{tabular}{cccccrc}
\hline\hline
&
\multicolumn{3}{c}{\sc equivalent-widths}
&
\multicolumn{3}{c}{\sc equivalent-width ratios}
\\
{\sc epoch} &
{${\OVI}$} &
{${\NV}$} &
{${\CIV}$} & 
{${\OVI}/{\CIV}$} &
{${\NV}/{\CIV}$} &
{${\NV}/{\OVI}$} \\
{(1)} &
{(2)} &
{(3)} &
{(4)} &
{(5)} &
{(6)} &
{(7)} \\
\hline
1 & 10.5$\pm$1.1 & 16.6$\pm$1.4 & 12.9$\pm$1.2 & 0.81$\pm$0.11 & 1.28$\pm$0.16 & 1.57$\pm$0.21 \\
2 & 17.6$\pm$0.9 & 27.8$\pm$1.0 & 29.0$\pm$1.4 & 0.61$\pm$0.04 & 0.96$\pm$0.06 & 1.58$\pm$0.10 \\
3 & 19.7$\pm$0.9 & 27.3$\pm$1.1 & 30.0$\pm$1.1 & 0.66$\pm$0.04 & 0.91$\pm$0.05 & 1.39$\pm$0.09 \\





%
\hline
\end{tabular}
\end{center}

Equivalent Widths (EW) and their errors are obtained from  equations~\ref{eq:EqWi} and \ref{eq:DEWi} of \S\ref{S:resu}
Col. (1): Observation Epoch. Cols. (2--4): Equivalent widths (in \AA) of \OVI, \NV\, and \CIV\ BALs. Cols. (5--7): Equivalent width ratios.

\end{minipage}
\end{table*}

The equivalent width (EW) of an absorption feature is defined as:
\begin{equation}
{\rm EW}=\int (1-f(\lambda))d\lambda,
\end{equation}
where $f(\lambda)$ is the normalized spectra. This is obtained by adding each pixel contribution through \citep{2002ApJ...574..643K}:
\begin{equation} \label{eq:EqWi}
{\rm EW}=\sum_i \left(1-\frac{F_i}{F_C} \right) B_i,
\end{equation}
where $i$ runs through every pixel, $F_i$ is the flux in the $i$th pixel, $B_i$ is the pixel width (in \AA), and $F_C$ is the continuum flux. Additionally the EW error is calculated by expanding the errors of the flux in each pixel and the error of the continua, thus obtaining:
\begin{equation}  \label{eq:DEWi}
\Delta {\rm EW}=\sqrt{\left( \frac{\Delta F_C}{F_C} \sum_i \frac{B_iF_i}{F_C} \right)^2+ \sum_i \left( \frac{B_i}{F_C} \Delta F_i \right)^2 }.
\end{equation}

For each of the three BAL features  (i.e., \OVI, \NV, and \CIV), valid pixels for EW calculations are those in the range of blueshifted velocities of $3000$ to $25000$~\kms, using as reference the wavelength of the \hbox{\OVI\ \lala1032}, \hbox{\NV\ \lala1239},  and \hbox{\CIV\ \lala1548} blue doublets, respectively.   Since the  \SiIV\ BAL is not observed, we obtain EW upper limits by multiplying by 2.3  the EW error (upper limit at the $\approx99$\% confidence level), which is obtained assuming $F_i=F_C$ in equation~\ref{eq:DEWi} and the wavelength of the blue doublet of the \hbox{\SiIV\ \lala1394}  as the reference.  These EW upper limits are $\sim 4$~\AA\ for every observation. 

For each epoch, the EWs and EW ratios of the  observed BALs in the \RF\ of \pg\ are presented in Table~\ref{tab:EqWi}.
From this table, we infer that there is no significant overall difference (within the errors) between the EWs obtained in Epochs~2 and 3. However, Epoch~1 shows weaker EWs in every observed BAL feature  when compared either with Epochs 2 or 3 (see Table~\ref{tab:EqWi}). We do not find any significant change in the EWs ratios between epochs. 
\begin{table*}
\begin{minipage}{175mm}
\begin{center}
\setlength{\tabcolsep}{1.6pt}
\caption{Spectral properties of BAL features \label{tab:BALs}}
\begin{tabular}{c @{~} @{\vline} @{~}  cccc @{~} @{\vline} @{~}  cccc @{~} @{\vline} @{~} cccc}
\hline\hline
 &
\multicolumn{4}{c}{\OVI\ (1031.9~\AA)} & 
\multicolumn{4}{c}{\NV\ (1238.8~\AA)} & 
\multicolumn{4}{c}{\CIV\ (1548.2~\AA)}  \\
{\sc epoch} &
{\vmin} & 
{\vmax} &
{\vmean} &
{$BI$} &
{\vmin} &
{\vmax} &
{\vmean} &
{$BI$} &
{\vmin} &
{\vmax} &
{\vmean} &
{$BI$} \\
{(1)} &
{(2)} &
{(3)} &
{(4)} &
{(5)} &
{(6)} &
{(7)} &
{(8)} &
{(9)} &
{(10)} &
{(11)} &
{(12)} &
{(13)} \\
\hline
1 & $ -5222$ & $-23136$ & $-14391$$\pm$2009 &  1594$\pm$242 & $-10094$ & $-24749$ & $-16527$$\pm$1960 & 2136$\pm$199 & $-10595$ & $-23336$ & $-16038$$\pm$1808 & 1046$\pm$105 \\
2 & $ -5326$ & $-25748$ & $-16110$$\pm$996 &  3900$\pm$226 & $ -6420$ & $-25039$ & $-16952$$\pm$799 & 5414$\pm$154 & $ -6381$ & $-25761$ & $-16520$$\pm$958 & 4626$\pm$213 \\
3 & $ -5127$ & $-23908$ & $-14974$$\pm$907 &  4472$\pm$210 & $ -6199$ & $-24767$ & $-16306$$\pm$844 & 5241$\pm$182 & $ -6417$ & $-27129$ & $-15951$$\pm$686 & 4688$\pm$173 \\
\hline
\end{tabular}
\end{center}

The units of (2--13) are \kms. 
Col. (1): Observation Epoch. Cols. (2--13): \vmin, \vmax, \vmean\ and BI for \OVI\ (Cols. 2--5), \NV\ (Cols. 6--9) and \OVI\ (Cols. 10--13).

\end{minipage}
\end{table*}

We obtain the balnicity index (hereafter BI) using the standard approach of \cite{1991ApJ...373...23W}, i.e., through:
\begin{equation}
BI=\int_{3000}^{25000}\left(1-\frac{f(-v)}{0.9}\right)Cdv,
\end{equation}
where $f(v)$ is the normalized spectrum at velocity $v$ in \kms\ in the rest frame.
The BIs starts to count from blueshifted velocities greater than 3000~\kms\ and $C$ is equal to zero unless the absorption depth in the normalized spectrum is greater than 10\% for an span of at least 2000~\kms.\footnote{The values of BI range from  0 (for \hbox{non-BAL} features) to  20000~\kms.}
As in the case of the EW, blueshifted velocities above 25000~\kms do not count for the BI calculations.
The velocity ranges where we find BAL absorption are \vmin\ and \vmax, and we assume that this absorption is present when the spectrum falls below 10\% of the continuum level.
For each ion, the calculations of BI, \vmin, and \vmax\ are obtained using boxcar smoothing over 13 pixels for the G140L spectra and 7 pixels for the G230L spectra (see Table~\ref{tab:BALs}). 
Since the BI index is a modification of the EW, we properly modify equations~\ref{eq:EqWi} and \ref{eq:DEWi} to obtain BIs with their errors. 
We also calculate a mean velocity of each BAL given by
\begin{equation}
\text{\vmean}=-\frac{\displaystyle\ \int_{3000}^{25000}  (1-f(-v))vdv}{\displaystyle\ \int_{3000}^{25000}  (1-f(-v))dv},
\end{equation}
where $f(v)$ is the normalized spectrum. The calculations of \vmean\ with their error were obtained adding each pixel contribution in a similar fashion as was done when obtaining the EWs.
As shown in Table~\ref{tab:BALs}, the BIs show similar tendencies as the EWs, and based on the \CIV\ line, \pg\ BIs range between 1000--4700~\kms. This range of BIs is consistent with a BI of $\approx2980$~\kms  obtained by \cite{2000ApJ...528..637B}  on a FOS/HST observation of \pg\ performed in 1992. Additionally, from Table~\ref{tab:BALs}, although some variations of \vmin\ and \vmax\ are observed, these seem to be  associated with changes in the optical depth and not with dynamical variations of the wind. This statement is confirmed by the mean velocity of the outflow ($v_{\rm mean}$), which does not show any significant variations either when we compare  between BAL features or observations (as seen in Table~\ref{tab:BALs}).

\section{Analysis and Discussion}

In this section, we analyze possble physical scenarios that could explain our results.  In \S\ref{S:clou} we introduce  a photoionization model that attempts to constrain basic physical properties of the UV outflow (e.g., the ionization state and hydrogen column density) based on the ranges of the EW and EW ratios presented in the last section. For this subsection, an optically thin BAL medium fully covering the central source is assumed. In \S\ref{S:tauf} we analyze an extension \S\ref{S:clou} by incorporating the covering fraction in the modeling of the UV wind.  Finally, in \S\ref{S:shie} we discuss the possible connection between the decrease in the soft X-ray emission and the increase in the BAL EWs found in Epochs~\hbox{1--3}.
\subsection{\textbf{\textsc{cloudy}} simulations} \label{S:clou}
\begin{figure}
	\includegraphics[width=8.4cm]{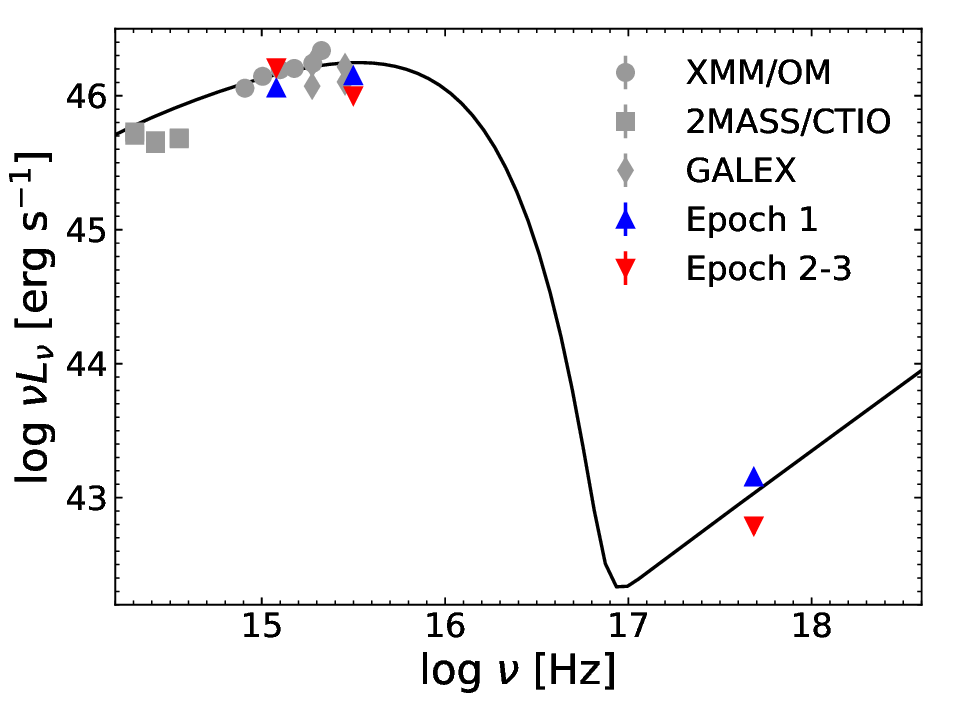}
	\centering
	\caption{SED used for our \cloudy\ simulations. For reference we plot  photometric data from:  \xmm\ Optical Monitor \citep[circles; from][]{2007A&A...474..431S}, 2MASS/CTIO \citep[squares; from][]{2006AJ....131.1163S}, \emph{GALEX} \citep[diamonds; from][]{2014AdSpR..53..900B}, and our observations (with triangles for Epoch 1 and inverted triangles obtained from the weighted mean \XR\ and UV fluxes from Epoch 2 and 3 in Tables~\ref{tab:flxx} and \ref{tab:fluv}). }
	\label{fig:PSED}
\end{figure}

Under the assumption that the absorbing gas producing the BAL features observed and analyzed in \S\ref{S:resu}  have a common origin, we use \cloudy\ \citep[version C17.vr;][]{2017RMxAA..53..385F}  to analyze the observed equivalent widths and ratios found in Table~\ref{tab:EqWi}. We perform \cloudy\ simulations assuming a point source AGN-type spectrum incident to an optically thin media.  
The spectral energy distribution (SED) of the point source spectrum is the {\sc agn} \cloudy\ model.\footnote{For more details on the {\sc agn} \cloudy\ model see \cite{1997ApJS..108..401K}.} 
The specific \cloudy\ command used was \texttt{AGN 5.5 -2.2 -0.5 0 }  in which 5.5 is the logarithm of the temperature (in kelvin) of the big blue bump, $-2.2$ is $\aox$, $-0.5$ is the low energy slope of the big blue bump $\alpha_{UV}$, and 0 is the X-ray slope $\alpha_x$ (where $\alpha_x=-\Gamma+1$). These parameters have been chosen to approximately fit the spectral energy distribution of past and current observations of the X-ray and UV spectra of \pg\ (see Fig.~\ref{fig:PSED}). The absorbing medium is given by a layer with total hydrogen column density of $\nh= 10^{13}~\cmsq$  fully covering the central source, with solar abundances and a fixed hydrogen density of $n_{\rm H} = 10^8~\cc$. The chosen value of $n_{\rm H}$ can be varied by at least three orders of magnitude without producing any noticeable effect on the results of our simulations \citep[see e.g.][]{1997ApJS..109..279H}. The value of \nh\ is chosen in order that the medium is ``thin enough"  so the most prominent lines that could form BALs have optical depths  $\tau_0$ in their centers which are significantly less than unity. 

The \cloudy\ simulations have been performed varying the values on the ionization parameter of the absorbing layer. The ionization parameter is the ratio of hydrogen ionizing photon density to hydrogen density, i.e.:
\begin{equation} \label{eq:ionp}
U= \frac{Q_{\rm H}} {4\pi r^2 n_{\rm H} c},
\end{equation}
where  $n_H$ is the total hydrogen density.  $Q_H$ is the rate of hydrogen ionizing photons emitted by the central object given by:
\begin{equation} \label{eq:Qhyd}
Q_{\rm H}= \int_{\nu_{Ry}}^{\infty} \frac{L_\nu d \nu}{h \nu},
\end{equation}
where $\nu_{Ry}$ is the frequency for photons with energies of 1 Rydberg ($Ry$).\footnote{Equation~\ref{eq:Qhyd} shows that  $Q_{\rm H}$ depends on the SED of the central object above $1\,Ry$.}
In the optically  thin regime for a fixed value of the ionization parameter and metallicity, the equivalent-width of an absorption line associated to the ion $j$ of column density $N_j$ should satisfy:
\begin{equation} \label{eq:self}
{\rm EW} \propto N_j \propto \nh.
\end{equation}

In particular,  $N_j$ can be obtained through the following expression:
\begin{equation} \label{eq:Ncol}
N_j \approx \frac{m_e c^2}{\pi e^2 f \lambda_0^2} {\rm EW},
\end{equation}
where $f$ is the oscillator strength and $\lambda_0$ is the laboratory wavelength of an absorption line associated to the ion $j$. Additionally, for Gaussian line profiles the optical depth at the line center of an absorption line produced by an ion $j$ is: 
\begin{equation}
\tau_0\approx \frac{ {\rm EW} \, c} {  \sqrt{\pi} \, b \, \lambda_0},
\end{equation}
where $b=\sqrt{2kT/m_j}$ is the most probable thermal velocity. If we assume that the BALs are the  added contribution of blue-shifted optically thin layers, our model can be properly scaled  to represent the BAL medium.  
From the curves presented in Figures~\ref{fig:ovin}--\ref{fig:civn}, if we assume $C_f=1$ we confirm that  the BAL medium maximum optical depth is $\tau_\lambda(\rm max) \lesssim 0.5$, and thus, we expect that our model can be used to estimate the ionization state and column density of the ions that produce the BALs.  By comparing these ion column densities with those obtained in our cloudy runs (with \lnh=13), through equations~\ref{eq:self} and \ref{eq:Ncol}, we can also extrapolate column densities of other ions or elements that do not necessarily  have observed spectral signatures (for example \nh). 
For partially covered outflows ($C_f<1$), which are commonly seen in BAL quasars \citep[see e.g.,][]{1999ApJ...524..566A, 2012ApJ...750..143D,2019ApJ...879...27L}, our model can be extended to provide useful insights into the ionization and column densities of the wind as described in \S\ref{S:tauf}.  
\begin{figure}
     \includegraphics[width=8.4cm]{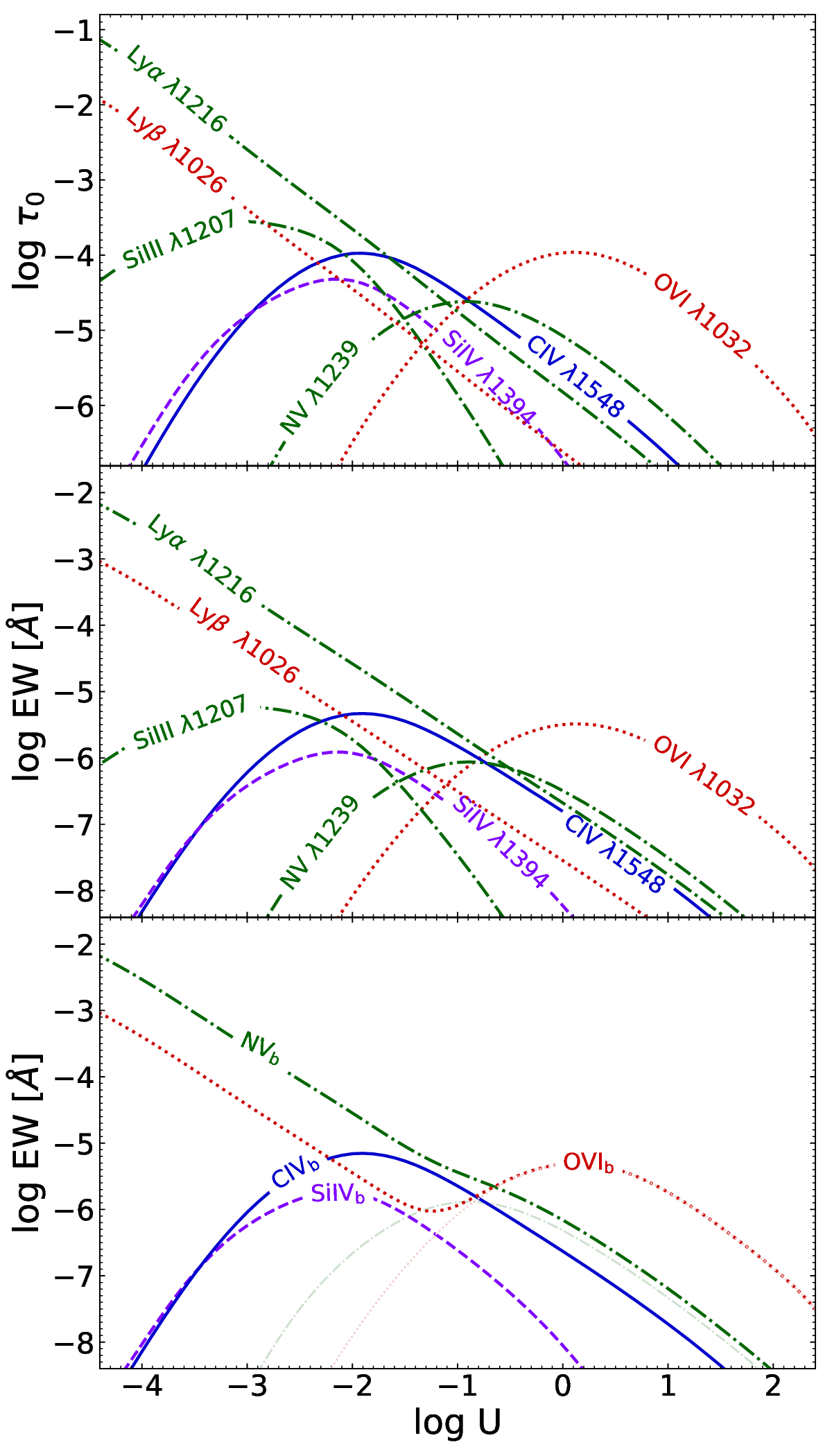}
	\centering
	\caption{{\bf Upper panel:} Logarithm of line-center optical depths versus logarithm of the ionization parameter. {\bf Middle panel:} Logarithm of the equivalent width versus logarithm of the ionization parameter.   {\bf Lower panel:} Logarithm of the combined equivalent width (EW) versus logarithm of the ionization parameter for line blending BAL regions. 
	In the upper and middle panels the lines marked correspond to the most prominent lines that are likely found in the BALs of the HST/STIS spectra.
	In all the panels the dotted, dash-dotted, dashed and full lines indicate lines belonging to the $\rm OVI_b$, $\rm NV_b$, $\rm SiIV_b$ and $\rm CIV_b$ blends, respectively (see main text for more details). In the lower panel  in a whiter tone the combined EW of the \OVI\ \lala1032,1038 (\hbox{dotted-line}), \NV\ \lala1239,1243 (\hbox{dash-dotted}) line doublets are marked. This figure is generated through \cloudy\ simulations of a point source AGN spectrum incident on an optically thin layer with a column density of $10^{13}\, \cmsq$.}
	\label{fig:tauW}
\end{figure}
\begin{figure}
	\includegraphics[width=8.4cm]{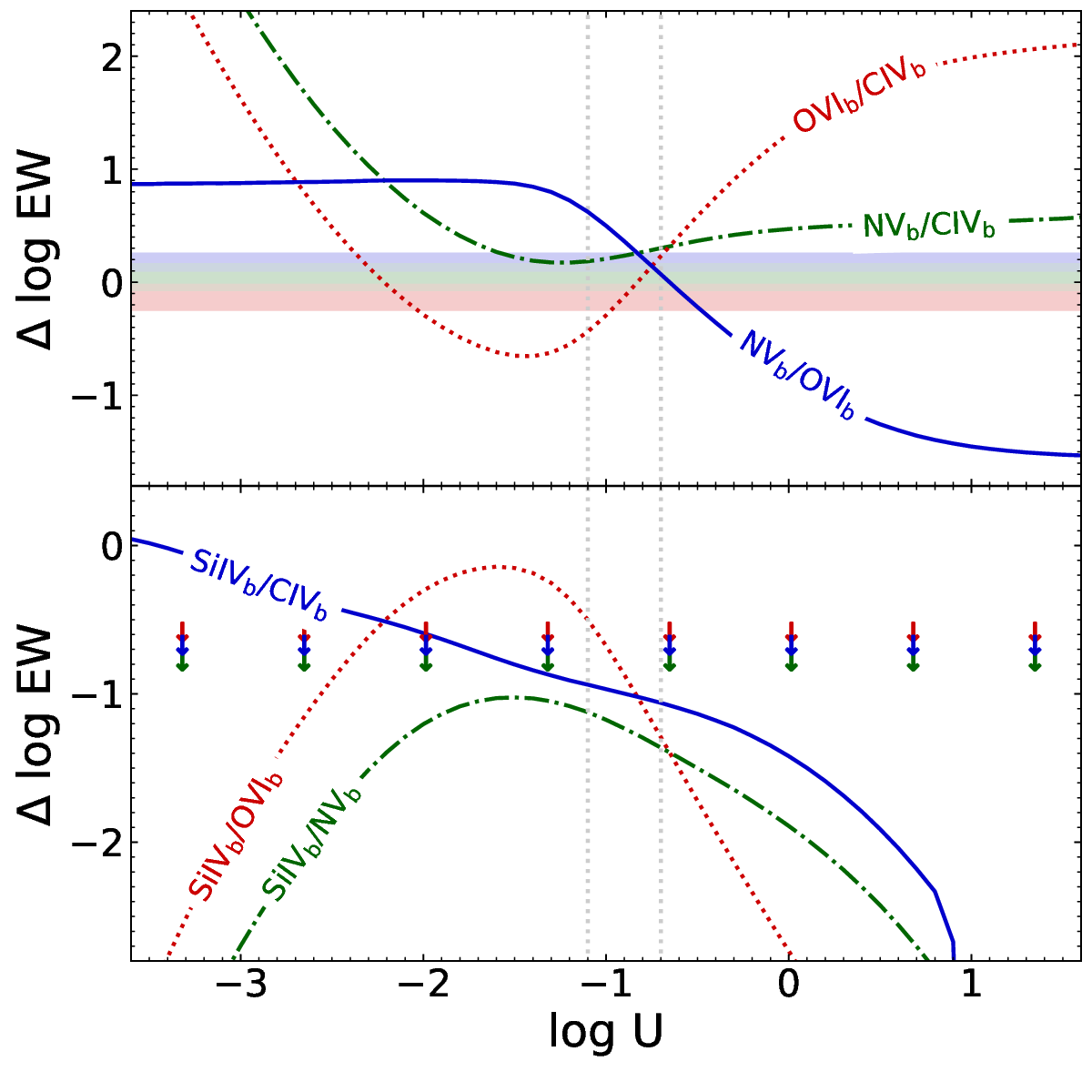}
	\centering
	\caption{ Logarithm of the equivalent width ratios of two line blends versus logarithm of the ionization parameter.  {\bf Upper panel:}  The dotted, \hbox{dash-dotted}, and full lines indicate the $\rm OVI_b/CIV_b$, $\rm NV_b/CIV_b$,  and $\rm NV_b/OVI_b$  EWs ratios, respectively. The red, green and blue horizontal bands represent the observed ranges of (in all Epochs) $\rm OVI_b/CIV_b$, $\rm NV_b/CIV_b$, and $\rm NV_b/OVI_b$ ratios (from Table~\ref{tab:EqWi}). {\bf Lower panel:} the dotted, \hbox{dash-dotted}, and full lines indicate the $\rm SiIV_b/OVI_b$, $\rm SiIV_b/NV_b$,  and $\rm SiIV_b/CIV_b$  EWs ratios respectively. The red, green and blue downward arrows mark the maximum   $\rm SiVb/OVI_b$, $\rm SiVb/NV_b$ and $\rm SiVb/CIV_b$ EW ratio upper limits in our observations.
		Both panels are generated through cloudy simulations of a point source AGN spectrum incident on an optically thin layer with a column density of $10^{13} \cmsq$. The vertical \hbox{dotted-line} band marks the expected $\log U$ range of the BAL medium  (see \S \ref{S:clou}).}
	\label{fig:EWbl}
\end{figure}

The absorption lines that we analyze are those that have significant signatures in the wavelength range of the \hst/STIS spectrum.  Additionally, these lines must cover the ionization states where the \CIV\ lines are significant. 
The most important of these lines are the \OVI\ \lala1032,1038, \NV\ \lala1239,1243, \SiIV\ \lala1394,1403 and \CIV\ \lala1548,1551 doublets. There are other lines that should have signatures in the spectral and ionization ranges of consideration; these are the \Lya\ \sla1216, \Lyb\ \sla1026, and the \SiIII\ \sla1207 lines. 
Given the wavelengths of these lines, we likely expect to find four blended sets of lines that could form BALs in our observations. The first, identified as the $\rm OVI_b$ blend is produced by the \Lyb\ line and the \OVI\ doublet. The second blend, identified as the $\rm NV_b$ blend is produced by the \Lya, \SiIII\ lines and the \NV\  doublet. 
Finally, the third and fourth, identified as the $\rm SiIV_b$ and $\rm CIV_b$ blends are produced by the \SiIV\ and \CIV\ doublets, respectively.

In Figure~\ref{fig:tauW} we show the optical depth at the line center $\tau_0$ (upper panel) and EW (middle panel) as a function of ionization parameter for the blue line of the \OVI, \SiIV, \NV\ and \CIV\ doublet with the \Lya, \Lyb, and the \SiIII\ lines. In the lower panel of Figure~\ref{fig:tauW} we show the logarithm of the added EWs of each line blend. 
To differentiate between line blends, in Figure~\ref{fig:tauW} each line in a particular blend is marked with a distinctive line style. The upper panel of this figure shows that all the lines have values of $\tau_0$ well below unity, thus confirming the fact that we are analyzing optically thin lines.
For our set of parameters, Figure~\ref{fig:tauW} indicates the ionization ranges where each line is contributing to potentially form a BAL. As we see in \S\ref{S:resu}, our observations show prominent $\CIV$ BAL signatures with no evidence of $\SiIV$ BAL features. Thus, if the BALs are produced by a medium with a narrow range of ionization states, we expect that $\log U \gtrsim -1$. 
 Additionally, we expect that $\log U \lesssim 1$ in order to be able to observe a significant \CIV\ BAL or \NV\ BAL as compared to the \OVI\ BAL . Therefore, in general we expect an outflow with $-1 \lesssim \log U \lesssim 1$, although a more refined ionization parameter range is obtained from analyzing the EW ratios as described in the following paragraphs.  In this ionization parameter range a comparison of the middle and lower panels in  Figure~\ref{fig:tauW} confirms that the most prominent lines producing BALs in our observations are the blue components of $\CIV$, $\OVI$ and $\NV$ doublets.

In  Figure~\ref{fig:EWbl}  the logarithms of the EW ratios of two different line blends as a function of $U$ are shown. 
The curves in the upper panel of Figure~\ref{fig:EWbl} correspond to ratios between the EWs of blends that could produce the observed BALs (i.e., OVIb, NVb, and CIVb blends). Additionally, the curves in the lower panel correspond to ratios between the EWs of the SiIVb blend and the OVIb, NVb, and CIVb blends, respectively.
In the upper panel of Figure~\ref{fig:EWbl} we have marked horizontal bands that correspond to the maximum and minimum EW ratios obtained from Table~\ref{tab:EqWi}. 
A simple observation of these bands indicates that there is no particular value of the ionization parameter that produces all the observed ratios in Table~\ref{tab:EqWi}.  
Additionally, the upper panel of Figure~\ref{fig:EWbl} shows that the observed values of the \NVb/\CIVb\ EW ratios are below the full-line curve.  
Thus, the discrepancies between our model and the observed EW ratios might be due to unknown complexity of the outflow, like variations from solar abundances of Oxygen, Nitrogen and Carbon  and/or gradients in the ionization state through the outflow. There could also be discrepancies due to the assumption of optical thinness  of our model, although we expect these should not be important as described in \S\ref{S:tauf}.
 
 The fact that the observed NVb/CIVb EW ratios are lower than expected from our model might be attributed to subsolar N/C abundance ratios as observed in quasar \hbox{HE 0141--3932} \citep{2005A&A...435...17R}. 
Since we cannot observe individual lines in the outflow and thus cannot estimate the relative abundances of O, N and C, in order to estimate a range for the ionization parameter of the wind, we add an error of 0.3 dex in the logarithm of the observed EW ratios. This error is close to what is expected from the dispersion of relative abundances of O, N and C in extragalactic \HII\ regions at nearly solar metallicities \citep[e.g.,][]{2009ApJ...700..654E, 2016ApJ...827..126B}.  
Based on the expanded error bands in the EW ratios, we obtain from the curves in Figure~\ref{fig:EWbl}  that $-1.2 \lesssim \log U \lesssim -0.7$,  $-1.8 \lesssim \log U \lesssim -0.2$, and $-1.1 \lesssim \log U \lesssim -0.5$ from the \OVI/\CIV, \NV/\CIV\ and \NV/\OVI\ ratios, respectively. Additionally, by comparing the observed \SiIV\ ratios upper limits with the curves in the lower panel of Figure~\ref{fig:EWbl}, we obtain $\log U \gtrsim -1.1$, and thus, we expect a range of the ionization outflow parameter $-1.1 \lesssim \log U \lesssim -0.7$.  Assuming that the ionization parameter is uniformly distributed in $[-1.1,-0.7]$ , and using the tables generated by our  \cloudy\ simulations, we obtain logarithms of relative abundance ratios with respect to hydrogen of $-4.21 \pm 0.14$,   $-4.40 \pm 0.05$, $-4.51 \pm 0.12$ for \OVI, \NV\ and \CIV\ respectively. From equations~\ref{eq:self} and \ref{eq:Ncol}, the abundance ratios obtained can be used to estimate (from the \OVI, \NV, and \CIV\ ion column densities) the BALs hydrogen column densities. 

Using the observed EW of each BAL in Table~\ref{tab:EqWi} we obtain an approximation of the column densities of the  \OVI, \NV, and \CIV\ ions through equation~\ref{eq:Ncol}. For this we assume that the EW of each observed BAL is produced by the added EWs of the  \OVI, \NV, and \CIV\ doublets respectively.\footnote{For a line doublet the ion column density $N_j$ can be obtained from: 
$$ N_j \approx \frac{m_e c^2}{\pi e^2(f_b \lambda_b^2+f_r \lambda_r^2)}{\rm EW},$$
where $b$ and $r$ stands for parameters of the blue and red doublet respectively.
}
Additionally, from the estimated relative abundances of ions with respect to hydrogen we estimate $\nh$ column densities from each ion, which are shown in Table~\ref{tab:EWNj}. As this table indicates, the medium that produces the BALs has  $\lnh \sim 20$ in Epoch~1 and its column density increases by $\approx0.3$~dex  from Epoch~1 to Epochs~2-3. 

\begin{table*}
\begin{minipage}{175mm}
\begin{center}
\caption{Observed column densities of  UV BALs  \label{tab:EWNj}}
\begin{tabular}{ccccccc}
\hline\hline
&
\multicolumn{3}{c}{\sc logarithm of ion column densities}
&
\multicolumn{3}{c}{\sc logarithm of h column densities}
\\
{\sc epoch} &
{\OVI} &
{\NV} &
{\CIV} & 
{\OVI} &
{\NV} &
{\CIV}  \\
{(1)} &
{(2)} &
{(3)} &
{(4)} &
{(5)} &
{(6)} &
{(7)} \\
\hline

1 & 15.75$\pm$0.04 & 15.72$\pm$0.04 & 15.33$\pm$0.04 & 19.96$\pm$0.15 & 20.11$\pm$0.06 & 19.84$\pm$0.13 \\
2 & 15.97$\pm$0.02 & 15.94$\pm$0.01 & 15.68$\pm$0.02 & 20.19$\pm$0.14 & 20.34$\pm$0.05 & 20.20$\pm$0.12 \\
3 & 16.02$\pm$0.02 & 15.93$\pm$0.02 & 15.69$\pm$0.01 & 20.23$\pm$0.14 & 20.33$\pm$0.05 & 20.21$\pm$0.12 \\




%
\hline
\end{tabular}
\end{center}

Col. (1): Observation Epoch. Cols. (2--4): Logarithm of ion column densities (in \cmsq) obtained using equation~\ref{eq:Ncol} from the equivalent widths (EWs) of the \OVI, \NV\, and \CIV\ BALs. Cols. (5--7):  Logarithm of hydrogen column densities as obtained from columns (2--4) and the ion abundances ratios with respect to hydrogen estimated by our thin layer \cloudy\ model (as described in \S \ref{S:clou}). 

\end{minipage}
\end{table*}

Equation \ref{eq:ionp} can be used to obtain $r_{UV}$, the distance from a central emitting source to the observed UV BAL wind. In view of our observations, we assume that $-1.1 \lesssim \log U \lesssim -0.7$ and $Q_H = 5.6 \times 10^{56}~{\rm s}^{-1}  $  from the SED obtained in Figure~\ref{fig:PSED}. Additionally, based on observations of BALs, mini-BALs and NALs,  the electron density of the outflow $n_e$ should lie between $3 \lesssim \log n_e \lesssim 10$ \citep[e.g.,][]{2001ApJ...548..609D,2011MNRAS.410.1957H, 2016ApJ...825...25M, 2019ApJ...876..105X}, with $n_{\rm H} \approx n_e$.\footnote{Assuming that the BALs have solar metallicities and totally ionized $n_e \approx 1.2\, \den$.} Based on these assumptions $r_{UV}$ is within $[10^3-10^7] R_S$ ($[0.1-1000]~{\rm pc}$),  where $R_S$ is obtained assuming a black hole mass of $M_{\rm BH}\approx 10^9 M_\odot$ \citep{2006ApJ...641..689V}. 

\subsection{Optical Depth Fits} \label{S:tauf}
In this section, we estimate the BAL ion column densities  by performing fits on the optical depths. This is done in order to analyze outflows that are not necessarily optically thin nor are totally covering the central source. 
The fits are performed by first transforming the normalized spectra to an optical depth profile as a function of wavelength ($\tau_\lambda$). The optical depth BAL profiles, assumed to be produced by the \OVI, \NV, and \CIV\  line doublets  (as in \S\ref{S:clou}), are deblended using the methodology described in \cite{1983ApJ...265...51J}. Using this approach the $\tau_\lambda$ profile is transformed to a single line with an oscillator strength of $f_*=f_b + f_r$ and a laboratory wavelength given by $\lambda_*=(f_b \lambda_b+f_r \lambda_r)/(f_b+f_r)$, where $f_b$ ($f_r$) is the oscillator strengths of the blue (red) doublet  and $\lambda_b$ ($\lambda_r$) is the laboratory wavelengths of the blue (red) doublet. 
Once deblended, each BAL profile is fitted as function of velocity using Gaussian profiles i.e.,
\begin{equation} \label{eq:gaus}
\tau_v=\tau_0 \exp(-((v-v_0)/b)^2), 
\end{equation}
where $\tau_0$ is an estimation of the maximum optical depth of the profile.
The ion column density of a single line absorption profile can be obtained from the optical depth velocity profile through the following expression:
\begin{equation} \label{eq:tauv}
N_j = \frac{m_e c}{\pi e^2   f_*  \lambda_*} \int \tau_v \, d  v ,
\end{equation}
 where  $\int \tau_v \, d  v=\sqrt{\pi} \,  \tau_0 \, b$ for a Gaussian profile.  
 Therefore, using equation~\ref{eq:tauv} we  show in Table~\ref{tab:Njta}  the column densities of the \OVI, \NV\ and \CIV\ ions using Gaussian fits on the optical depth profiles of each observation. Additionally, in  Table~\ref{tab:Njta}, we show the estimated hydrogen column densities  using the relative ion abundance ratios with respect to the hydrogen obtained in \S\ref{S:clou}.  For $C_f=1$, the results in Table~\ref{tab:Njta}  (as expected) are very similar to those of Table~\ref{tab:EWNj}. For $C_f<1$, which is the case of an absorber partially covering the emitting source, the observed and true optical depths of an absorption line are obtained from:
\begin{equation} \label{eq:cfta}
 \exp(-\tau_\lambda^{\rm obs})=(1-C_f)+C_f \exp(-\tau_\lambda^{\rm true}).
 \end{equation}  
Since $\tau_\lambda^{\rm true}$ grows monotonically with column density, in absence of constraints on the column density, the assumption of $C_f=1$ results in column densities that are lower limits.  In Table~\ref{tab:Njta} we have performed again Gaussian fits to the deblended ionic profiles  but now with $C_f=0.8$ and 0.6. The case of $C_f=0.6$ is a rough estimate of the minimum value expected from Epochs 2 and 3.\footnote{It is expected that $C_f \geq 1-f_{\rm min}$ were $f_{\rm min}$ is the minimum normalized residual flux from an absorption trough \citep[e.g.,][]{2003ARA&A..41..117C}.} 
From Table~\ref{tab:Njta}, as expected, the column densities and maximum optical depths ($\tau_0$) vary approximately inversely  proportional with $C_f$. Additionally, from this table, we conclude that for a given BAL profile if $C_f$ is close to its expected minimum then $\tau_0\gtrsim 1$.  For this particular case, i.e. of saturated line profiles, our model might not give reliable estimates of $U$ and/or $\nh$.  

The approach used to describe the UV BAL winds in \S\ref{S:clou} and this section is mainly based on constraining the medium ionization parameter and describing ion column densities of each BAL. This methodology, although simple should give us reliable constraints on the expected column densities and the ionization state of the wind. To clarify this better, we generate two \cloudy\ runs motivated by the optical depth Gaussian fits in the case of $C_f=0.8$.
Both runs  are from  a layer with solar abundance, with ionization parameter at the illuminated face  $\log U =-0.8$,  and turbulent velocity  $v_{\rm turb}=6500~\kms$. The difference in the runs are the chosen values of column densities  which are $\lnh=20.1$ and $\lnh=20.5$ respectively. The chosen value of $U$ is within the expected for the UV BAL medium,  $v_{\rm turb}$ is close to the average value of the estimated velocity dispersions of the Gaussian fits ($b$ parameter in equation~\ref{eq:gaus}), and the values of $\lnh$ are close to the average values estimated for Epoch~1 and  Epochs~\hbox{2-3} (for $C_f=0.8$ in Table~\ref{tab:Njta}). 
The absorption spectra from these \cloudy\ runs are  then corrected for partial covering (using equation~\ref{eq:cfta} with $C_f=0.8$) and  blue-shifted with a blue-shifted velocity of 16000~\kms. This velocity is close to the average Gaussian central velocities  ($v_o$ in equation~\ref{eq:gaus}) obtained from the fits. The resulting  absorption profiles are shown in Figures~\hbox{\ref{fig:ovin}--\ref{fig:civn}}.  By comparing these curves with the observed profiles, and given the simplicity of our model assumptions, we find a good agreement with the observations. Note that similar absorption profiles to those presented in  Figures~\hbox{\ref{fig:ovin}--\ref{fig:civn}} can be obtained for the case of $C_f=1$ and $C_f=0.6$.
The use of more sophisticated photoionization models \citep[e.g.,][]{2018ApJ...866....7L, 2019ApJ...879...27L} requires a clearer definition of the continuum at \RF\ wavelengths $\lesssim 1200$~\AA, and additional evidence of spectral features (not found in this work) that could provide tighter constraints in the covering fraction and/or the chemical abundance of the BALs medium.

\begin{table*}
\begin{minipage}{175mm}
\begin{center}
\caption{Column densities and maximum optical depths from deblended Gaussian fits to BAL optical depth profiles \label{tab:Njta}}
\begin{tabular}{ccccccccccc}
\hline\hline
&
\multicolumn{3}{c}{\sc logarithm of ion column densities}
&
\multicolumn{3}{c}{\sc logarithm of h column densities}
&
\multicolumn{3}{c}{\sc  Fitted values of $\tau_0$}
\\
{\sc epoch} &
{\OVI} &
{\NV} &
{\CIV} & 
{\OVI} &
{\NV} &
{\CIV} & 
{\OVI} &
{\NV} &
{\CIV}  \\
{(1)} &
{(2)} &
{(3)} &
{(4)} &
{(5)} &
{(6)} &
{(7)} &
{(8)} &
{(9)} &
{(10)} \\

\hline
 & \multicolumn{9}{c}{\sc covering fraction  Cf= 1.00} \\
\hline
1 & 15.83$\pm$0.04 & 15.72$\pm$0.04 & 15.38$\pm$0.02 & 20.04$\pm$0.15 & 20.12$\pm$0.06 & 19.89$\pm$0.12 & 0.23$\pm$0.01 &  0.32$\pm$0.02 & 0.24$\pm$0.01 \\
2 & 16.11$\pm$0.05 & 16.02$\pm$0.03 & 15.79$\pm$0.03 & 20.32$\pm$0.15 & 20.42$\pm$0.06 & 20.30$\pm$0.12 & 0.38$\pm$0.02 &  0.62$\pm$0.03 & 0.56$\pm$0.02 \\
3 & 16.12$\pm$0.04 & 16.00$\pm$0.03 & 15.81$\pm$0.03 & 20.33$\pm$0.15 & 20.40$\pm$0.06 & 20.32$\pm$0.12 & 0.46$\pm$0.03 &  0.61$\pm$0.03 & 0.53$\pm$0.02 \\
\hline
 & \multicolumn{9}{c}{\sc covering fraction  Cf= 0.80} \\
\hline
1 & 15.93$\pm$0.05 & 15.84$\pm$0.04 & 15.49$\pm$0.02 & 20.14$\pm$0.15 & 20.24$\pm$0.07 & 20.00$\pm$0.12 & 0.30$\pm$0.02 &  0.42$\pm$0.03 & 0.31$\pm$0.01 \\
2 & 16.23$\pm$0.05 & 16.15$\pm$0.03 & 15.92$\pm$0.03 & 20.44$\pm$0.15 & 20.55$\pm$0.06 & 20.43$\pm$0.12 & 0.51$\pm$0.03 &  0.85$\pm$0.05 & 0.76$\pm$0.03 \\
3 & 16.24$\pm$0.05 & 16.14$\pm$0.03 & 15.93$\pm$0.03 & 20.45$\pm$0.15 & 20.54$\pm$0.06 & 20.44$\pm$0.12 & 0.62$\pm$0.04 &  0.84$\pm$0.05 & 0.72$\pm$0.03 \\
\hline
 & \multicolumn{9}{c}{\sc covering fraction  Cf= 0.60} \\
\hline
1 & 16.08$\pm$0.05 & 15.99$\pm$0.04 & 15.63$\pm$0.02 & 20.29$\pm$0.15 & 20.39$\pm$0.07 & 20.14$\pm$0.12 & 0.42$\pm$0.03 &  0.61$\pm$0.04 & 0.44$\pm$0.01 \\
2 & 16.40$\pm$0.06 & 16.37$\pm$0.04 & 16.10$\pm$0.03 & 20.61$\pm$0.15 & 20.77$\pm$0.07 & 20.61$\pm$0.12 & 0.77$\pm$0.05 &  1.43$\pm$0.10 & 1.22$\pm$0.05 \\
3 & 16.42$\pm$0.05 & 16.35$\pm$0.04 & 16.12$\pm$0.03 & 20.63$\pm$0.15 & 20.75$\pm$0.07 & 20.63$\pm$0.12 & 0.99$\pm$0.07 &  1.44$\pm$0.11 & 1.20$\pm$0.07 \\
\hline
\end{tabular}
\end{center}

Col. (1): Observation Epoch. Cols. (2--4): Logarithm of ion column densities (in \cmsq) obtained using equation~\ref{eq:tauv} on optical dephs obtained through Gaussian fits of the deblended optical depth profiles of the \OVI, \NV\, and \CIV\ BALs. Cols. (5--7): Logarithm of hydrogen column densities as obtained from columns (2--4) and the ion abundances ratios with respect to hydrogen as described in \S \ref{S:clou}. Cols. (8--10) maximum optical depth from Gaussian fits of the deblended optical depth profiles of the \OVI, \NV\, and \CIV\ BALs.

\end{minipage}
\end{table*}

\subsection{Is there a Connection between variability of the BALs and the X-rays?} \label{S:shie}
With a limited sample of observations, we are not able to establish any conclusive argument regarding an \XR\ wind connection in \pg. However, our analysis has revealed some trends  that could shed light on what to expect for future observations.  The most relevant tendency shown in our observations is  that a weakening in the soft \XR\ flux could be associated with an increase of the EW of the BALs (see Figures~\hbox{\ref{fig:flxx}--\ref{fig:civn}}). The mean velocities of the BALs did not change significantly in our observations; thus, we do not find evidence that the X-rays are connected with the acceleration mechanism of the UV wind. 

If the BALs are produced from an optically thin medium, fully covering the central source, the increase in the EW could be due to a growth in the column density of the wind, which doubled between Epoch~1 and Epochs~2-3  (as seen in  \S\ref{S:clou} and \S\ref{S:tauf}). Therefore, a weakening in the \XRs\ could be associated with a more massive outflow. 
In the case of a partially covered outflow,  the EW increase of the BALs might be also linked to a growth in the covering fraction of the wind. As noted in \S\ref{S:resu} the increase in EW of the BALs is also associated with a redder UV spectra as measured by \aeuv\ and \afuv\ (see Table~\ref{tab:fluv}). This  change of the UV spectra may be related to a decrease of the wind inclination with respect to the line of sight  \citep{2013MNRAS.432.1525B}. A  redder UV spectrum might also be a consequence of an increase in the column density (mass) of the outflow. The combined effect of a more massive and/or less inclined wind might produce an  increase in the covering fraction and/or the column density. Note that if the UV wind is partially covering the central source, then the BAL column densities of $\lnh \sim 20$ obtained in \S\ref{S:clou} are interpreted as lower limits (see Table~\ref{tab:Njta}).

Epoch~2 corresponds to the weakest state ever recorded in X-rays. The variability in the  X-rays  between Epoch~1 and Epoch~\hbox{2-3} could be interpreted as an increase in the column density of the shield (see Table~\ref{tab:fpar}). Unfortunatelly, our new X-ray observations have low S/N, and thus, they are not sufficient to give any conclusive statement. 
From \cite{2012ApJ...759...42S}, when comparing past \XR\ observations of \pg, Epoch~1 is close to average \XR\ brightness. Additionally, in the  \asca\ observation performed in 1999 \citep[hereafter identified as Epoch~0;][]{2001ApJ...546..795G}, which corresponds to the brightest \XR\ observation registered, the \SB\ \XR\ flux is  approximately 10 times greater than that of Epoch~1 (see Figure~\ref{fig:flxx}). For the brightest \XR\ observation, a simple absorbed power law model with  $\lnh\sim22$ gives a plausible fit to the data  \citep{2001ApJ...546..795G, 2012ApJ...759...42S}.  
In conclusion, a shielding layer with  a varying \lnh\ from $\sim\,22.0$ to $\sim\,23.0$ might at least explain in part the strong X-ray variability that this source has.  
The premise of a shield layer with varying column density  is also reinforced by the observed values of the 2~keV \RF\ monochromatic fluxes of \pg. To demonstrate this, we assume that the intrinsic  (unabsorbed)  monochromatic flux at 2~keV  is $f_{\rm 2keV} \approx  11 \times 10^{-31} \flux$, which is obtained assuming that the 2500~\AA\  luminosity is given by the weighted average of \hbox{Epochs~1--3} and the 2~keV luminosity is from equation~3 of \cite{2007ApJ...665.1004J}. 
We further assume that every \XR\  observation is a product of an APL model with a varying column density and a fixed photon index ($\Gamma=1.9$). Therefore, with $f_{\rm 2 keV}\approx 8 \times 10^{-31} \flux$ for the brightest \XR\ observation  \citep[Epoch~0; see Table~9 from][]{2012ApJ...759...42S}, and the monochromatic 2~keV \RF\ fluxes in Table~\ref{tab:flxx},  we estimate $\lnh \approx 22.0$, 22.8 and 22.9 for Epoch~0, Epoch~1 and Epoch~2-3, respectively. 
As described in \S\ref{S:XRan}, we expect the X-ray absorption of \pg\ is more complex than that obtained from an APL model. However, under the assumption that the intrinsic X-ray emission does not vary significantly, the APL model should provide good approximations to the column densities that produce the observed changes in soft X-ray emission.

In the wind-shield scenario \citep{1995ApJ...451..498M, 2007ASPC..373..305G}, the UV wind is attributed to the outermost portions of the wind while the X-ray shield to the innermost zones of the wind. As stated in \S\ref{S:clou}, the UV wind should be at distances from the central source of $r_{\rm UV} \gtrsim 1000 R_S$. Given the \XR\ variability found between Epochs~2 and 3 of $\sim\!5$~months in the \RF, and assuming a dynamic shield, a coherent light crossing time argument gives a distance to the central source of $r_{\rm X} \lesssim 1000 R_S$  \citep[where $M_{\rm BH}\approx 10^9 M_\odot$ is assumed;][]{2006ApJ...641..689V}.  Thus,  we expect that the absorption in X-rays is produced by a different medium than that producing the UV BAL features. Based on other observations of BAL quasars, we might expect \XR\ variability associated to shield signatures on time scales shorter than 5 months in the \RF\  \citep[e.g.,][]{2009ApJ...706..644C, 2020ApJ...895...37R}.  
Under a wind-shield scenario interpretation, a change in the inclination and/or mass of the wind could produce an increment in the column density and/or the covering fraction.  
Associated to these changes, we also expect a nearly contemporaneous effect in the inner zone of the wind, as observed with a likely increase in the column density of the shield after Epoch~1. In our observations, we do not detect significant changes in the EW ratios of the observed BALs, and thus, we cannot establish if the increase in EWs is associated with changes in the ionization state of the wind. If we assume a shielded wind-scenario in an observation where the X-ray emission flux rises to a close to brightest state, we likely expect less massive winds and changes in the incident SED that could indicate variation in the ionization state of the wind (i.e., variability in the EW ratios).

Based on this analysis, it would be fundamental to perform a multi-wavelength monitoring campaign of this source  to confirm or reject the trends found in this work.  At this time, it is not  possible to elucidate if a complex absorption or simply an intermittent \XR\ source is producing the strong variability that this source shows. This dilemma might be resolved with long exposure \nustar\ observations, in order to further constrain the model associated with the fast and dramatic \XR\ variability of \pg.  As described in this work, albeit our results are preliminary, a shielded wind model seems to give a plausible explanation to our \hst/\chandra\ spectra.

\section{Summary and Conclusions}

In this work, we analyze three sets of \chandra\ observations each with a contemporaneous \hst\ STIS spectra (Epochs~\hbox{1--3}) of \pg. Epoch~1 was performed in 2002, while Epochs~2 and 3 were performed in 2014--2015 and separated by approximately nine months.  Our main conclusions are the following:

\begin{enumerate}[labelwidth=0.3cm,leftmargin=0.4cm,align=left,label= \arabic*.]
\item There is significant \XR\ spectral hardening between Epoch~1 and Epoch~3. The soft \XR\ flux ($\lesssim\!2$~keV) does not change significantly between  Epochs~2  and 3. Additionally, the \SB\ flux is reduced by approximately one third when Epoch~1 is compared either with Epoch~2 or Epoch~3. 

\item The \hst\  UV spectra exhibit BAL features that are likely associated with the \OVI\ \lala1032,1038, \NV\ \lala1239,1243, and \CIV\ \lala1548,1551 doublets. These observations, in general show that Epoch~1 is significantly distinct from either Epoch~2 or Epoch~3.  Additionally, Epoch~3 has only minor changes in relation to Epoch~2. When Epoch~1 is compared with Epochs~2-3, the EWs of the BALs are found to increase and the UV spectral slope becomes redder. 

\item Under the assumption that BALs are produced by an optically thin medium fully covering the source, we estimate hydrogen column densities of \mbox{$\sim 10^{20} \cmsq$} for Epoch~1 that increase by a factor of approximately two for Epoch 2-3.  If the medium is partially covered ($C_f<1$), the outflow column densities obtained assuming $C_f=1$ become lower limits and the covering fraction might be also contributing to the changes of EWs. In conclusion, the increasing EWs can be explained by an increase in column densities and/or covering fractions. These changes accompanied by a redder UV spectrum might be indicating  a decrease in the inclination of the outflow along the line of sight and/or an increase in the wind mass. The EW ratios of the \CIV, \NV, and \OVI\ BALs do not change significantly between observation Epochs, implying that the BALs medium ionization parameter is stable.  

\item The decrease in the soft ($\lesssim 2$~keV) \XR\ flux, that is produced between Epoch~1 and Epoch~2-3, is possibly associated with an increase in the absorption of the shield. This growth is close to contemporaneous  with increments of the BALs EWs. 

\end{enumerate}

The analysis of the X-ray/UV spectra of \pg, provide support for the wind-shield scenario model of BAL winds.   A future multiwavelength long-term monitoring campaign of this source in combination with a more sophisticated UV spectral  analysis \citep[e.g., with SimBAL;][]{2018ApJ...866....7L} would be of interest  to confirm or reject the conclusions obtained here.

\section{Acknowledgements}

Support for this work was provided by the National Aeronautics and Space Administration through Chandra Award Number GO5-16119X issued by the Chandra X-ray Observatory Center, which is operated by the Smithsonian Astrophysical Observatory for and on behalf of the National Aeronautics Space Administration under contract \hbox{NAS8-03060}. Support for Program number \hbox{HST-GO-13948.001-A} was provided by NASA through a grant from the Space Telescope Science Institute, which is operated by the Association of Universities for Research in Astronomy, Incorporated, under NASA contract \hbox{NAS5-26555}. 
We acknowledge support from ANID grants CATA-Basal AFB-170002 (FEB), FONDECYT Regular 1190818 (FEB), 1200495 (FEB) and Millennium Science Initiative Program - ICN12\_009 (FEB).
This work was partially supported by JSPS KAKENHI Grant Number 21H01126. SCG acknowledges support from the Natural Science and Engineering Research Council of Canada.

\section{Data Availability}

The scientific results reported in this article are based on publicly available observations made by the Chandra \XR\ Observatory and the Hubble Space Telescope (\hst). The list of \chandra\  observations IDs (ObsIDs) are: 3011, 17148, and 17553.\footnote{Downloadable from \url{https://cxc.harvard.edu/cda/}} Additionally, the  \hst\ observations are from Proposal IDs 13948 and 9277.\footnote{Downloadable from: \url{https://archive.stsci.edu/}}

\bibliography{msbib}

\end{document}